\documentclass{amsart}
\usepackage{mathrsfs}
\usepackage{amssymb}
\usepackage{verbatim}
\usepackage{enumitem}
\usepackage{color}
\usepackage{xcolor}

\newtheorem{theorem}{Theorem}

\newtheorem{proposition}{Proposition}[section]

\theoremstyle{definition}
\newtheorem{definition}{Definition}[section]
\newtheorem{notation}{Notation}
\newtheorem{example}{Example}[section]

\newtheorem{exercise}{Exercise}[section]

\newtheorem{remark}{Remark}[section]

\numberwithin{equation}{section}

\def\Xint#1{\mathchoice
{\XXint\displaystyle\textstyle{#1}}%
{\XXint\textstyle\scriptstyle{#1}}%
{\XXint\scriptstyle\scriptscriptstyle{#1}}%
{\XXint\scriptscriptstyle\scriptscriptstyle{#1}}%
\!\int}
\def\XXint#1#2#3{{\setbox0=\hbox{$#1{#2#3}{\int}$}
\vcenter{\hbox{$#2#3$}}\kern-.5\wd0}}

\def\dashint{\Xint-}

\definecolor{light-grey}{gray}{0.80}

\begin{document}

\title[Impurity models and products of random matrices]{Impurity models and products of random matrices}
\author{alain Comtet}
\address{LPTMS\\
        Universit\'e Paris 11\\
        91400, Orsay, France}
\address{Sorbonne Universit\'es, UPMC, Universit\'{e} Paris 06 \\ 70005, Paris, France}
\email{alain.comtet@u-psud.fr}
\author{Yves Tourigny}
\address{School of Mathematics\\
        University of Bristol\\
        Bristol BS8 1TW, United Kingdom}
\email{y.tourigny@bristol.ac.uk}



\thanks{It is our pleasure to thank Christophe Texier for kindly providing us with some of the figures used, and for making a number of suggestions which improved an earlier draft of these notes}


\begin{abstract}
This is an extended version of lectures given at the {\em Summer School on Stochastic Processes and Random Matrices}, held at 
the \'{E}cole de Physique, Les Houches, in July 2015.
The aim is to introduce the reader to the theory of one-dimensional disordered systems and  products of random matrices, confined to the $2 \times 2$ case. The notion of {\em impurity model}--- that is, a system in which the interactions are highly localised---  links the two themes and enables their study by elementary mathematical tools.
After discussing the spectral theory of some impurity models, we state and illustrate Furstenberg's theorem, which gives sufficient conditions for the exponential growth of a product of independent, identically-distributed matrices.
\end{abstract} 

\maketitle

\section{Introduction}
\label{introductionSection}
\subsection{Product of matrices}
\label{ProductSubsection}
Consider a sequence
$$
A_1,\, A_2,\, A_3,\, \ldots
$$
of invertible $d \times d$ matrices drawn independently from some distribution, say $\mu$. In these notes,
we shall be interested in the large-$n$ behaviour of the product
$$
\Pi_n := A_n \, A_{n-1} \cdots A_1\,.
$$
More precisely, we shall investigate conditions under which the product {\em grows exponentially} with $n$ and seek to compute
the growth rate.

The solution of this problem is of great interest in the physics of disordered systems. Such systems are often modelled in terms of linear difference or differential equations with random coefficients and, in order to understand the behaviour of the system, one
must study the spectral problem
associated with the equation. For the models we shall consider, the general solution of the equation can be expressed in terms of the product $\Pi_n$ for a suitable choice of the $A_n$.

\subsection{Disordered systems}
\label{disorderSubsection}
The existence of widely separated scales is a remarkable feature of the physical world. The fact that the corresponding degrees of freedom can--- to a very good approximation--- decouple makes it possible to construct effective theories of condensed matter physics, and even of particle physics. The same crucial simplification occurs also in the physics of disordered systems: one can distinguish between fast variables which evolve very quickly and  slow variables which, in a real experiment, are frozen in a specific configuration. It is then legitimate to treat the slow variables as (static but) random variables, distributed according to a prescribed probability law. Such systems are usually modelled using the concept of {\em random operator}.

Consider for instance the quantum mechanics of an electron interacting with a collection of scatterers, and think of these
scatterers as ``impurities'' in an otherwise homogeneous medium.
This problem is modelled by the  Hamiltonian
$$
{H} =\frac{{\mathbf p}^2}{2m}+\sum_{j=1}^{n}V({\mathbf r}-{\mathbf r}_j)
$$
where the random parameters include, among others, the positions ${\mathbf r}_j$ of the impurities and their number $n$. We could for instance assume that the positions of the scatterers are independent and distributed uniformly in space and consider the thermodynamic limit for a fixed density.
Although this model looks very simple, its analytical treatment presents substantial difficulties; 
even basic quantities of physical interest, such as the large-volume distribution of eigenvalues, can seldom  be computed exactly. 

In the one-dimensional case, however, the problem is more
tractable: much progress can be made by considering  {\em initial-value problems.}
Techniques pioneered by
Dyson (1953) and Schmidt (1957)  have been developed that lead in a few instances to explicit formulae. Schmidt's justification for bothering with the one-dimensional case was the``hope that it gives in some respects a qualitatively correct description of real three-dimensional [systems]''. Dyson, on the other hand, felt that
``interest for working on one-dimensional problems is merely that they are fun''. Our own view is that the value of exact calculations is in revealing possible connections between objects which, at first sight, may have seemed unrelated. 

The primary purpose of these lectures is to elaborate the connection between the Dyson--Schmidt methodology, applied 
to a certain class of disordered systems, and one of this school's themes, namely
products of random matrices. Historically, it may be said that the desire to understand the behaviour of disordered systems 
provided one of the main motivations 
for the development, by Furstenberg (1963) and others, of the theory of products of random matrices. The classic reference on the interplay
between the two subjects is the book by Bougerol and Lacroix (1985). That book is divided into two parts: the first 
part gives a rigorous account of Furstenberg's theorem on the growth of products of matrices; the second part applies that theory
to products obtained from Anderson's model, which uses a ``tight-binding''  (i.e. a finite-difference) 
approximation  of the Schr\"odinger equation with a random potential.  
Roughly speaking, the programme there is to apply Furstenberg's theory to some disordered systems and in so doing deduce localisation properties
of the disordered states.

\subsection{Outline}
\label{outlineSubsection}
In the present lectures, we revisit these connections, but the flow of ideas is in the opposite direction: we begin by considering a class of disordered
systems which use the notion of {\em random point interaction}. We call these disordered systems {\em impurity models}. We explain how, via the well-known technique of separation of variables, these models lead to products of $2 \times 2$ matrices, and we undertake the study of their spectral properties;
it will be seen that these are intimately linked with the growth of the solutions.
We also briefly touch upon a number of related topics, including the scattering problem for a disordered sample.
In applying and developing these ideas in the context of our impurity models, we encounter many of the objects that feature in Furstenberg's work.
Our approach thus provides some physically-motivated insights into the abstract mathematics of the
Furstenberg theory.

\subsection{Recommended reading}
\label{readingSubsection}
The lectures constitute an extensive development of some of the ideas presented in the papers 
by Comtet, Texier and Tourigny (2010; 2011; 2013) and Comtet, Luck, Texier and Tourigny (2013); from the mathematical point of view, they are more or less self-contained, in the sense that anyone who is familiar with the basic concepts of probability, differential equations, complex variables, and group theory should be able to learn something from them.  

We make no attempt to provide a systematic survey of the literature, but merely refer to those papers and books that we have personally found helpful in developing our own understanding of the subject. For an alternative introduction to disordered systems, the reader is encouraged to consult Pastur's survey (Pastur 1973), or the more elaborate account in Lifshits {\em et al.} (1988). The first chapter of Luck's excellent but insufficiently known monograph is also highly recommended (Luck 1992).

The material leading to Furstenberg's theorem is based on his fundamental paper (Furstenberg 1963), and on the later monographs by Bougerol and Lacroix (1985) and Carmona and Lacroix (1990).

\section{Some impurity models}
\label{impuritySection}

This section presents some examples of impurity models and explains how products of matrices arise naturally from them.

\subsection{The vibrating string and Dyson's random chain model}
\label{dysonSubsection}
Consider an inextensible string of unit tension, tied at the ends of the interval $[0,L]$. Denote by
$M(x)$ the total mass of the string segment corresponding to the interval $[0,x)$ and by $y(x,t)$ the vertical displacement
of the string from its equilibrium position above the point $x$ at time $t$. The Lagrangian associated with this mechanical system
is
$$
\frac{1}{2} \int_0^L \left [ M'(x) \left ( \frac{\partial y(x,t)}{\partial t} \right )^2 - \left ( \frac{\partial y(x,t)}{\partial x} \right )^2\right ]\,{\rm d} x\,.
$$
It follows from Hamilton's principle of least action--- see
for instance Simmons (1972)--- that $y(x,t)$ obeys the {\em wave equation}
$$
M'(x) \frac{\partial^2 y(x,t)}{\partial t^2} -  \frac{\partial^2 y(x,t)}{\partial x^2} = 0\,, \;\; 0 < x < L\,,\; t > 0\,.
$$
This equation admits separable solutions of the form
$$
y(x,t) = \psi(x) \,{\rm e}^{ {\rm i} \omega t}
$$
provided that $\omega$ is a characteristic frequency of the string, i.e. a number such that
there exists a non-trivial solution of the two-point boundary-value problem
\begin{equation}
\psi'' (x) + \omega^2 M'(x) \,\psi(x) = 0\,,\;\;0 < x < L\,, \quad \psi(0) = \psi(L) = 0\,.
\label{stringEquation}
\end{equation}

For an ``ideal'' string with a  uniform distribution, the characteristic frequencies may be calculated exactly (see Exercise \ref{weylExercise}). More realistically, however, the manufacturing
process may cause variations in the thickness of the string, or the material of which the string is made may contain defects. The statistical study of the effect of such imperfections on the characteristic
frequencies, in the large-$L$ limit, was initiated  by Dyson (1953). He
considered
the particular case of a string consisting of point masses, i.e.
\begin{equation}
M'(x) = \sum_{j=1}^\infty m_j \,\delta \left ( x-x_j \right )
\label{discreteString}
\end{equation}
where the masses $m_j > 0$ and the positions
\begin{equation}
0 =:  x_0 < x_1 < x_2  < \cdots 
\label{partition}
\end{equation}
are {\em random}.  It follows from the definition of the Dirac delta
that the differential equation in (\ref{stringEquation})
admits solutions that are continuous and piecewise linear with respect to the partition (\ref{partition}). More precisely,
after setting
$$
\lambda := \omega^2
$$
we have the recurrence relations
\begin{equation}
\psi'( x_{j+1}-) - \psi'(x_j-) + \lambda m_j \psi (x_j) = 0 
\label{firstDiscreteStringEquation}
\end{equation}
for $j=1,\,2,\,\cdots$, 
and
\begin{equation}
\psi'(x_{j+1}-) = \frac{\psi(x_{j+1})-\psi(x_j)}{\ell_j}
\label{secondDiscreteStringEquation}
\end{equation}
for $j=0,\,1,\,\cdots$, where
\begin{equation}
\ell_j := x_{j+1}-x_j\,.
\label{spacing}
\end{equation}
Expressed in matrix form, these relations become
\begin{equation}
\begin{pmatrix}
\psi'(x_{j+1}-) \\
\psi(x_{j+1})
\end{pmatrix}
= A_j 
\begin{pmatrix}
\psi'(x_{j}-) \\
\psi(x_{j})
\end{pmatrix}
\label{discreteStringEquation}
\end{equation}
where
\begin{equation}
A_j := \begin{cases}
 \begin{pmatrix}
1 & 0 \\
\ell_0 & 1
\end{pmatrix} & \text{if $j=0$} \\
 & \\
 \begin{pmatrix}
1 & 0 \\
\ell_j & 1
\end{pmatrix}
\begin{pmatrix}
1 & -\lambda m_j \\
0 & 1
\end{pmatrix} & \text{otherwise} 
\end{cases}\,.
\label{discreteStringMatrix}
\end{equation}
Hence we may express the solution of the {\em initial-value problem} for the string equation as
\begin{equation}
\begin{pmatrix}
\psi'(x_{n+1}-) \\
\psi(x_{n+1})
\end{pmatrix}
= \Pi_n
A_0 \begin{pmatrix}
\psi'(0-) \\
\psi(0) 
\end{pmatrix}\,.
\label{discreteStringSolution}
\end{equation}
where $\Pi_n$ is the product of
matrices
\begin{equation}
\Pi_n := A_{n} \cdots A_2 A_1\,. 
\label{productOfMatrices}
\end{equation}

We shall return in \S \ref{spectralSection} to the spectral problem for this string.

\begin{remark}
Our notation differs from Dyson's; his string equation is
$$
K_j \left ( x_{j+1}-x_{j} \right ) + K_{j-1} \left ( x_{j-1}-x_j \right ) = -m_j \omega^2 x_j\,.
$$
In his notation, the $x_j$ are the positions of particles coupled together by springs that obey Hooke's law, and $K_j$ is the elastic modulus
of the spring between the $j$th and $(j+1)$th particles. The correspondence between 
this and Equations (\ref{firstDiscreteStringEquation}-\ref{secondDiscreteStringEquation}) is
$$
x_j \sim \psi (x_j) \;\;\text{and}\;\; K_j \sim 1/\ell_j\,.
$$
\label{dysonRemark}
\end{remark}

\begin{center}
\linethickness{2mm}
\color{light-grey}{\line(1,0){360}}
\end{center}
\begin{exercise}
\label{stringSolutionExercise}
For $\omega^2 = \lambda$, let $\psi(\cdot,\lambda)$ be the particular solution of Equation (\ref{stringEquation}) such that $\psi(0,\lambda) = \cos \alpha$ and $\psi'(0-,\lambda) = \sin \alpha$.
Show by induction on $n$ that, for the mass density (\ref{discreteString}), $\psi'(x_{n+1}-,\lambda)$ and $\psi(x_{n+1},\lambda)$ are polynomials of degree $n$ in $(-\lambda)$ and that 
the leading
coefficient of $\psi(x_{n+1},\lambda)$ is
$$
c_n := \left ( \cos \alpha + \ell_0 \sin \alpha \right ) \prod_{j=1}^n \left ( \ell_j m_j \right )\,.
$$
\end{exercise}

\begin{exercise}
\label{generalisedStringExercise}
Consider the slightly more general mass distribution
$$
M'(x) = \mu + \sum_{j=1}^\infty m_j \,\delta \left ( x-x_j \right )\,,\;\;\mu \ge 0\,.
$$
Show that the solution of the string equation (\ref{stringEquation}) is again of the form (\ref{discreteStringSolution})
if the $A_j$ are appropriately modified.
\end{exercise}
\begin{center}
\linethickness{2mm}
\color{light-grey}{\line(1,0){360}}
\end{center}

\subsection{The Frisch--Lloyd model}
\label{frischLloydSubsection}
The fundamental equation of quantum mechanics is the time-dependent Schr\"odinger equation; see for instance Texier (2011).
For a single particle on the positive half-line, the method of separation of variables leads to the following time-independent version:
\begin{equation}
- \psi ''(x) + V(x) \psi (x) = E \psi (x)\,, \quad x > 0,
\label{schroedingerEquation}
\end{equation}
where $V$ is the potential function and $E$ is the energy of the particle. 

Consider a potential of the form
\begin{equation}
V(x) = \sum_{j=1}^\infty v_j \,\delta(x-x_j)\,.
\label{potential}
\end{equation}
One interpretation for this choice of potential is as follows: there are impurities located at the points $x_j$ of the
partition (\ref{partition}),
and $v_j$ is the ``coupling constant'' of the interaction at $x_j$ (Frisch and Lloyd 1960). 
For such a potential, the general solution of Equation (\ref{schroedingerEquation})
may again be constructed in a piecewise fashion: For $x_{j} < x < x_{j+1}$ and $E=k^2$ with
$k>0$,
$$
\begin{pmatrix}
\psi' ( x ) \\
 \psi (x )
\end{pmatrix}
= \begin{pmatrix}
\cos \left [ k (x-x_j) \right ] & -k \sin \left [ k (x-x_j) \right ] \\
\sin  \left [ k (x-x_j) \right ]/k & \cos \left [ k (x-x_j) \right ] 
\end{pmatrix} \,\begin{pmatrix}
1 & v_j \\
0 & 1
\end{pmatrix}
\,
\begin{pmatrix}
\psi' (x_j-) \\
 \psi (x_j-)
\end{pmatrix}\,.
$$
Iterating, we deduce
$$
\begin{pmatrix}
\psi' ( x_{n+1}- ) \\
 \psi (x_{n+1}- )
\end{pmatrix}
= \Pi_n A_0
\begin{pmatrix}
\psi' (0-)  \\
\psi(0) 
\end{pmatrix}
$$
where the matrices in the product $\Pi_n$, defined by Equation (\ref{productOfMatrices}), are now given by
\begin{equation}
A_j := 
\begin{pmatrix}
\cos (k\ell_{j}) & -k \sin (k \ell_{j}) \\
\sin ( k \ell_{j})/k & \cos (k \ell_{j})
\end{pmatrix} 
\begin{pmatrix}
1 & v_j \\
0 & 1
\end{pmatrix} 
\label{frischLloydMatrix}
\end{equation}
and
$$
A_0 := 
\begin{pmatrix}
\cos (k\ell_{0}) & -k \sin (k \ell_{0}) \\
\sin ( k \ell_{0})/k & \cos (k \ell_{0})
\end{pmatrix} 
 \,.
$$

It is useful to comment briefly on the {\em deterministic case}
$$
A_j = A := 
\begin{pmatrix}
\cos (k \ell) & -k \sin (k \ell) \\
\sin (k \ell)/k & \cos (k \ell)
\end{pmatrix}
\begin{pmatrix}
1 & v \\
0 & 1
\end{pmatrix}  
\;\; \text{for every $j \in {\mathbb N}$}\,.
$$
The Frisch--Lloyd model then reduces to the famous Kronig--Penney model which was introduced long ago to analyse the band structure of crystalline materials (Kronig and Penney 1931). 
The asymptotic behaviour of this product is readily determined 
by examining the eigenvalues and eigenvectors of $A$. Since $A$ has unit determinant, the reciprocal of an eigenvalue of $A$
is also an eigenvalue. In particular, if $\left | \text{Tr} A \right |>2$ then $A$ has two distinct
real eigenvalues--- one of which must exceed unity--- and the product grows exponentially with $n$.
On the other hand, if  $\left | \text{Tr} A \right | < 2$, then the eigenvalues form a complex conjugate pair on the unit circle and there is no growth.
In physical terms, this means that, for a periodic system without disorder, the allowed values of the energy are given by those ranges of $k$ for which the inequality  
$$ 
|\cos (k\ell) +  \sin (k \ell) \,v/(2k)| < 1
$$ 
is satisfied. For other values of the energy, there are no traveling or Bloch-like solutions, so that forbidden gaps in the energy spectrum are formed. 

\begin{center}
\linethickness{2mm}
\color{light-grey}{\line(1,0){360}}
\end{center}
\begin{exercise}
Show that, if one considers instead the case $E=-k^2 < 0$, $k >0$,
then the general solution of the Frisch--Lloyd model is in terms of a product of matrices where
\begin{equation}
A_j := 
\begin{pmatrix}
\text{ch} (k\ell_j) & k\,\text{sh} (k \ell_j) \\
\text{sh} ( k \ell_j )/k & \text{ch} (k \ell_j)
\end{pmatrix} 
\begin{pmatrix}
1 & v_j \\
0 & 1
\end{pmatrix} 
\,.
\label{negativeEnergyFrischLloydMatrix}
\end{equation}
\label{frischLloydExercise1}
\end{exercise}
\begin{center}
\linethickness{2mm}
\color{light-grey}{\line(1,0){360}}
\end{center}

\subsection{The Anderson model}
\label{andersonSubsection}
For the mathematical description of the {\em localisation phenomenon}, P. W. Anderson used the following
approximation of the stationary Schr\"{o}dinger equation (Anderson 1958):
$$
-\psi_{n+1} + V_n  \psi_n - \psi_{n-1} = E \psi_n\,, \;\; n \in {\mathbb N}\,.
$$
This is in fact the model that is usually studied in introductions to the theory of disordered systems (Luck 1992).
The concepts that underly its mathematical treatment are analogous to those used for our impurity models, but we shall not consider them
in what follows (see, however, \S \ref{furstenbergSection}). The principal benefit of restricting our attention to  impurity models
is that, since they are described by differential (as opposed to difference) equations, one
can make full use of the tools of differential calculus.

\subsection{Further motivations}
\label{motivationSubsection}

Products of random matrices are encountered not only in quantum impurity models but also in the classical statistical physics of disordered systems. 
Consider for instance a one-dimensional Ising chain consisting of $n$ spins embedded in an external inhomogeneous magnetic field. The Hamiltonian can be written in the form
$$
{H} (\sigma) := - J \sum_{j=1}^n  \sigma_j\sigma_{j+1}-\sum_{j=1}^n h_j \sigma_j\,.
$$
In this expression, $J$ is the coupling constant  and $h_j$ is the local magnetic field at site $j$ which is linearly coupled to 
the spin $\sigma_j=\pm 1$ there. For a positive (negative) coupling constant, the interaction is ferromagnetic (anti-ferromagnetic); this favours the alignment or anti-alignment of neighbouring spins. In terms of the slow/fast dichotomy mentioned in the introduction, we assume here that the spin variables are the fast variables, and that they are in thermal equilibrium in a specific frozen configuration $\{h_j\} $. The object is then to evaluate the canonical partition function of the frozen system at temperature $\beta=1/(kT)$:
$$
{Z}_n=\sum_{\sigma} {\rm e}^{-\beta {H}(\sigma) }\,.
$$
Physical observables are in principle extracted  by taking the thermodynamic limit $n \rightarrow \infty$.  A remarkable feature of disordered systems is  that certain extensive quantities such as the free energy 
$$
-\frac{kT}{n} \ln {Z}_n 
$$  
converge as $n \rightarrow \infty$ to a non-random limit with probability 1. Such quantities are called {\em self-averaging}; this means that a typical realisation of a large enough system gives access to the thermodynamic quantities. 
The self-averaging property of the free energy is a consequence of 
the central limit theorem for products of random matrices, and 
it enables one to express the free energy of the random Ising chain in terms of the growth rate of a certain product.

The easiest way of calculating $Z_n$ is to use the transfer matrix technique. The partition functions $Z_{j}^{\pm}$, conditioned on the last spin $\sigma_j=1$ or $\sigma_j=-1$, obey the following recurrence relation
\begin{equation}
\notag
\begin{pmatrix}
Z_j^{+} \\
Z_j^{-}
\end{pmatrix}
= \begin{pmatrix}
{\rm e}^{\beta(J+h_j)} & {\rm e}^{\beta(-J+h_j)} \\
{\rm e}^{\beta(-J-h_j)} & {\rm e}^{\beta(J-h_j)}
\end{pmatrix} 
\,
\begin{pmatrix}
Z_{j-1}^{+} \\
Z_{j-1}^{-}
\end{pmatrix}\,.
\end{equation}
For a chain of $n$ spins with periodic boundary conditions the partition function takes the form
$$
Z_n= \text{Tr} \,\Pi_n
$$
where the matrices in the product (\ref{productOfMatrices}) are 
$$
A_j = \begin{pmatrix}
{\rm e}^{\beta(J+h_{j})} & {\rm e}^{\beta(-J+h_{j})} \\
{\rm e}^{\beta(-J-h_{j})} & {\rm e}^{\beta(J-h_{j})}
\end{pmatrix} \,.
$$

\section{The spectral problem}
\label{spectralSection}
In the previous section, we used the method of separation of variables to reduce the solution of some time-dependent
models to particular instances of the following ``master'' system:
\begin{equation}
-\psi'' + Q' \psi = \lambda M' \psi\,, \quad x > 0\,
\label{masterEquation}
\end{equation}
subject to the condition
\begin{equation}
\sin \alpha \,\psi(0) - \cos \alpha \,\psi'(0) =0\,.
\label{leftCondition}
\end{equation}
Here $\{ Q(x): x \ge 0\}$ and $\{M(x): x \ge 0\}$,  with $M$ non-decreasing, are two processes,
$\lambda$ is the spectral parameter, 
and $\alpha \in [0,\pi)$ is some fixed parameter independent of $\lambda$. 
We have chosen to introduce $\alpha$ in order to make explicit the dependence of the spectral quantities on the condition at
$x=0$. The case $\alpha=0$ corresponds to a Neumann condition, and the case $\alpha=\pi/2$ to a Dirichlet condition.

Every student of physics and mathematics is familiar with the next stage of the method of separation of variables: the object is to find all the values of the spectral parameter for which a non-trivial solution exists, and then to expand
the solution of the {\em time-dependent problem} in terms of these ``eigenfunctions''. In the simplest cases, such as the 
heat and wave equations with constant coefficients, the eigenfunctions are trigonometric functions;
the expansion is then the  familiar Fourier series or transform, and there exist corresponding inversion
formulae which permit the recovery of the solution from its Fourier coefficients or transform.

Our purpose in this section is to outline the extension of these concepts to the more complicated equation
(\ref{masterEquation}). In the general case, there is of course no hope of having explicit formulae for the ``eigenvalues''
and ``eigenfunctions'' of the problem. Nevertheless, we shall see that the necessary information 
for the construction of Fourier-like
expansions is contained in a so-called {\em spectral measure}, which is itself accessible via an important object
called the {\em Weyl coefficient} of the spectral problem. This beautiful theory, 
developed by Weyl and Titchmarsh in the deterministic case, brings out the important part played by the asymptotic behaviour of the solutions for large $x$. Our account is based on Coddington \& Levinson (1955), Chapter 9. Some care must be taken when this theory is applied to Dyson's impurity model, because in that case the coefficient $M'$ is not smooth. We shall indicate briefly the necessary adjustments that have been worked out by M. G. Kre\u{\i}n and his school (Kac and Kre\u{\i}n 1974).

It will be useful to denote by $\psi(\cdot,\lambda)$ and
$\varphi(\cdot,\lambda)$ the particular solutions of the differential equation that satisfy the initial conditions
\begin{equation}
\psi(0,\lambda) = -\varphi'(0,\lambda) = \cos \alpha \;\;\text{and}\;\;\psi'(0,\lambda) = \varphi (0,\lambda) = \sin \alpha\,.
\label{fundamentalSolutions}
\end{equation}

In technical terms, the spectral problem is {\em singular}
because the independent variable $x$ runs over an infinite interval. In order to understand the essential features of the singular case, we begin by considering a truncated version: For $L >0$,
\begin{equation}
-\psi'' + Q' \psi = \lambda M' \psi\,, \quad 0< x < L\,,
\label{truncatedMasterEquation}
\end{equation}
subject to the conditions
\begin{equation}
\sin \alpha\, \psi(0) - \cos \alpha \, \psi'(0) = 0 \;\;\text{and}\;\;\psi'(L-) = z \,\psi(L) \,.
\label{boundaryCondition}
\end{equation}
The parameter $z$ that is used to specify the boundary at $x=L$ can take any real value (independent of the spectral parameter $\lambda$), as well as the ``value'' $\infty$; the boundary condition at $x=L$ is then interpreted as $\psi(L)=0$, i.e. Dirichlet's
condition. We shall see that it is very instructive to consider the entire range of this parameter.

To discuss the spectral problem, we introduce the operator
\begin{equation}
{\mathscr L} := \frac{{\rm d} x}{{\rm d} M} \left [ - \frac{{\rm d}^2}{{\rm d} x^2} + \frac{{\rm d} Q}{{\rm d}x}  \right ]
\label{operator}
\end{equation}
associated with the
problem (\ref{truncatedMasterEquation}-\ref{boundaryCondition}). When $M$ is smooth, 
${\mathscr L}$ is a classical differential operator. In particular, we then have {\em Green's identity}:
\begin{multline}
\int_0^L {\mathscr L} u (x)\,\overline{v(x)}\, {\rm d} M (x) - \int_0^L u(x) \,\overline{{\mathscr L} v (x)}\, {\rm d} M (x) \\
= \left [ u(x)\,\overline{v'(x)} - u'(x) \,\overline{v(x)} \right ]  \Bigl |_{0+}^{L-}
\label{greenIdentity}
\end{multline}
for every sufficiently smooth complex-valued functions $u$ and $v$ defined on $[0,L]$. This key identity implies,
among other things, that the Wronskian of the particular solutions $\varphi(\cdot,\lambda)$ and $\psi(\cdot, \lambda)$
is identically equal to $1$. More importantly, it also follows that ${\mathscr L}$ is {\em self-adjoint} with respect to the inner product
\begin{equation}
(u,\,v) := \int_0^L u (x) \,\overline{v(x)}\,{\rm d} M(x) 
\label{innerProduct}
\end{equation}
in a suitable space of functions that satisfy the boundary conditions (\ref{boundaryCondition}).
We associate with this inner product the norm $\| \cdot \| := \sqrt{(\cdot,\cdot)}$.

For Dyson's model, however, $M$ is merely piecewise constant
with respect to the partition (\ref{partition})
and the interpretation of Equation (\ref{operator}) requires some clarification. A thorough discussion may be found in Kac \& Kre\u{\i}n (1974). For our immediate purposes, it will be sufficient to remark that, if we accept 
(\ref{operator}) and proceed formally, then
$$
\int_0^L {\mathscr L} u(x) \,\overline{v(x)} \,{\rm d} M(x) = \int_0^L \overline{v(x)}\, \left [ -{\rm d} u'(x) + u(x) \,{\rm d} Q(x)  \right ]\,.
$$
For instance, if $u$ is continuous and piecewise linear with respect to the partition (\ref{partition}), the meaning of the integral on the right-hand side is clear; by using the ``integration by parts'' formula
$$
\int_0^L \overline{v(x)}\, {\rm d} u'(x) = u'(x)\,\overline{v(x)} \Bigl |_{0+}^{L-} - \int_0^L u'(x)\,\overline{v'(x)}\,{\rm d} x
$$
it may be shown that Green's identity remains valid for Dyson's model.

\subsection{The spectral measure}
\label{measureSubsection}

The eigenvalues of ${\mathscr L}$ are the zeroes of the function
$$
\lambda \mapsto \psi'(L-,\lambda) - z \,\psi(L,\lambda)\,.
$$
For $z \in {\mathbb R} \cup \{\infty\}$,
they are
real, simple and may be ordered:
$$
-\infty < \lambda_1 < \lambda_2 < \cdots 
$$
The eigenfunction corresponding to $\lambda_j$ is a multiple of $\psi (x,\lambda_j)$. Furthermore, every
function $f$ whose norm is finite may be expressed as a ``Fourier'' series:
$$
f(x) = \sum_{j} f_j \psi(x,\lambda_j)\,, \quad f_j := \frac{\left ( f, \psi(\cdot,\lambda_j) \right )}{\| \psi (\cdot,\lambda_j ) \|^2}\,.
$$
This expansion may be expressed neatly in the form
\begin{equation}
f(x) = \int_{\mathbb R} \widehat{f}_L (\lambda) \,\psi(x,\lambda)\, {\rm d} \sigma_L (\lambda) 
\label{FiniteExpansion}
\end{equation}
where $\widehat{f}_L$ is the ``transform''
\begin{equation}
\widehat{f}_L (\lambda) := \int_0^L f(x) \,\psi (x,\lambda)\,{\rm d} M(x)
\label{finiteTransfrom}
\end{equation}
and
\begin{equation}
\sigma_L ' (\lambda) = \sum_{j} \| \psi(\cdot,\lambda_j ) \|^{-2} \delta ( \lambda-\lambda_j)\,.
\label{finiteMeasure}
\end{equation}
The measure $\sigma_L$ is called the {\em spectral measure} associated with the truncated problem. 

\subsection{The integrated density of states per unit length}
\label{integratedSubsection}

The spectral measure is a complicated object. For the models we have in mind, it is the eigenvalues that
may be measured experimentally and, for this reason, physicists are often more interested in
the function
\begin{equation}
N_L (\lambda) := \frac{\# \left \{ n \in {\mathbb N} :\,\lambda_n < \lambda  \right \}}{L}
\label{dysonCountingMeasure}
\end{equation}
which simply counts the number of eigenvalues per unit length.
Thus, $N_L$ retains only
``half the 
information'' contained in the spectral measure; the norm
of the eigenfunctions $\psi(\cdot,\lambda_j)$ has been lost.

The measure $N_L$ may have a weak limit as $L \rightarrow \infty$; that is,
there may be a measure $N$ such that, for every smooth function $\eta$ with compact support in ${\mathbb R}$,
there holds
$$
\int_{-\infty}^\infty \eta(\lambda) \,{\rm d} N_L (\lambda ) \xrightarrow[L \rightarrow \infty]{} 
\int_{-\infty}^\infty \eta(\lambda) \,{\rm d} N (\lambda )\,.
$$
We call this limit measure, if it exists, the {\em integrated density of states} per unit length.

\subsection{The Weyl coefficient}
\label{weylSubsection}
The spectral characteristics of the truncated problem may be accessed by considering another particular solution
of the differential equation (\ref{masterEquation}); it is defined as the linear combination
\begin{equation}
\chi (x,\lambda) = \varphi(x,\lambda) + w_L(\lambda) \psi(x,\lambda)
\label{linearCombination}
\end{equation}
where the coefficient $w_L(\lambda)$ is chosen so that
\begin{equation}
\chi_L' (L-,\lambda) = z \,\chi_L (L,\lambda)\,.
\label{recessiveSolution}
\end{equation}
It is then readily seen that
\begin{equation}
w_L (\lambda) =  -\frac{\varphi'(L-,\lambda)-z\,\varphi(L,\lambda)}{\psi'(L-,\lambda)-z\,\psi(L,\lambda)}\,.
\label{weylCoefficient}
\end{equation}

The function $\lambda \mapsto w_L(\lambda)$ is called the {\em Weyl coefficient} associated with the truncated spectral problem. It turns out that the spectral measure  $\sigma_L$ may be recovered from it.

\begin{notation}
For a function $F$ of the complex variable $x+ {\rm i} y$, we write
$$
F(x \pm  {\rm i} 0 ) := \lim_{0 < \varepsilon \rightarrow 0} F(x \pm {\rm i} \varepsilon)\,.
$$
\end{notation}

\begin{theorem}
\begin{equation*}
\sigma_L(\lambda) - \sigma_L(\lambda_0) = 
\frac{1}{\pi}  \int_{\lambda_0}^\lambda \text{\em d} \zeta  \,\text{\em Im} \,w_L(\zeta+\text{\em i} 0 )
\end{equation*}
at points of continuity $\lambda$ and $\lambda_0$ of $\sigma_L$.
\label{coddingtonLevinsonTheorem}
\end{theorem}
A proof may be found in Coddington \& Levinson (1955), Chapter 9, \S 3.

\begin{center}
\linethickness{2mm}
\color{light-grey}{\line(1,0){360}}
\end{center}
\begin{exercise}
For the particular case $\alpha =\pi/2$, $Q =0$ and $M' = 1$:
\begin{enumerate}[label=(\alph*)]
\item Show that
$$
w_L(\lambda) = \frac{z \,\sqrt{\lambda}\,\cot \left ( \sqrt{\lambda} L \right ) - \lambda}{\sqrt{\lambda}\,\cot \left ( \sqrt{\lambda} L\right )-z}\,.
$$
The Weyl coefficient is thus a meromorphic function of $\lambda$, with poles at the eigenvalues 
$\lambda_n$. 

\item For the case $z = \infty$, compute the residue at $\lambda_n$. Thus show
 by comparison with Equation (\ref{finiteMeasure}) that Theorem \ref{coddingtonLevinsonTheorem} does indeed hold in this case.
{\em Help:} In order to evaluate the integral of the imaginary part of Weyl's coefficient, use Cauchy's Theorem with the rectangular contour
of height $2 \varepsilon > 0$ centered on the interval $0 < x < \lambda$. Then let $\varepsilon \rightarrow 0$ \,.

\item Show that $N_L$ has a weak limit as $L \rightarrow \infty$ given by
$$
N(\lambda) = 
\frac{1}{\pi}\sqrt{\lambda}\,,\;\;\text{for $\lambda > 0$}\,.
$$
\end{enumerate}
\label{weylExercise}
\end{exercise}

\begin{exercise}
Show that
$$
w_L(\lambda) = \frac{\frac{\chi_L'(0,\lambda)}{\chi_L (0,\lambda)}\,\sin \alpha+\cos \alpha}{\sin \alpha - \frac{\chi_L'(0,\lambda)}{\chi_L (0,\lambda)} \,\cos \alpha}\,.
$$
\label{weylRiccatiExercise}
\end{exercise}
\begin{center}
\linethickness{2mm}
\color{light-grey}{\line(1,0){360}}
\end{center}

\subsection{The Riccati equation}
\label{riccatiSubsection}
Let us pause for a moment in order to draw attention to the very simple form taken by the Weyl coefficient for our impurity models.
Set
$$
Z (x) = \frac{\chi_L'(x,\lambda)}{\chi_L(x,\lambda)}\,.
$$
The foregoing exercise shows that the Weyl coefficient may be expressed in terms of $Z(0)$. In particular,
we have
$$
w_L(\lambda ) = Z(0) = \chi_L'(0,\lambda)  \;\;\text{for $\alpha = \pi/2$}\,.
$$
On the other hand, since $\chi_L(\cdot,\lambda)$ solves the truncated equation (\ref{truncatedMasterEquation}), 
$Z$ is the
particular solution of the Riccati equation
\begin{equation}
Z' =  -Z^2 + Q' - \lambda M'\,, \;\; x > 0\,,
\label{riccatiEquation}
\end{equation}
that satisfies the condition $Z(L-) = z$.

For our impurity models, we can construct this particular solution by proceeding as follows: we associate with 
$$
A = \begin{pmatrix}
a & b \\
c & d
\end{pmatrix} \in \text{SL} (2, {\mathbb R})
$$
the linear fractional transformation ${\mathcal A} : \, {\mathbb C} \cup \{ \infty \} \rightarrow {\mathbb C} \cup \{ \infty \}$
defined by
\begin{equation}
{\mathcal A} (z) = \begin{cases}
a/c & \text{if $z = \infty$} \\
\frac{a z + b}{c z + d} & \text{otherwise}
\end{cases}\,.
\label{linearFractionalTransformation}
\end{equation}
These linear fractional transformations form a group for the operation of composition.

\begin{center}
\linethickness{2mm}
\color{light-grey}{\line(1,0){360}}
\end{center}
\begin{exercise}
Verify the formulae
\begin{enumerate}[label=(\alph*)]
\item 
$$
A = \begin{pmatrix}
1 & v \\
0 & 1
\end{pmatrix}
\;\;\implies \;\;{\mathcal A}(z) = v + z\,.
$$
\item
$$
A = \begin{pmatrix}
1 & 0 \\
\ell & 1
\end{pmatrix}
\;\;\implies \;\;{\mathcal A}(z) = \cfrac{1}{\ell + \cfrac{1}{z}}\,.
$$
\item 
$$
A = \begin{pmatrix}
1 & 0 \\
\ell & 1
\end{pmatrix}
\begin{pmatrix}
1 & -\lambda m \\
0 & 1
\end{pmatrix}
\;\;\implies \;\; {\mathcal A} (z) = \cfrac{1}{\ell + \cfrac{1}{-\lambda m + z}}\,.
$$

\item 
$$
A = \begin{pmatrix}
\cos(k \ell) & -k \sin(k \ell) \\
\sin(k \ell)/k & \cos(k \ell)
\end{pmatrix}
\;\;\implies {\mathcal A}(z) = \cfrac{1}{\tau + \cfrac{1+k^2 \tau^2}{-k^2 \tau + z}}
$$
where $\tau = \tan(k \ell)/k$.

\item
$$
A = \begin{pmatrix}
\cos(k \ell) & -k \sin(k \ell) \\
\sin(k \ell)/k & \cos(k \ell)
\end{pmatrix}
\begin{pmatrix}
1 & v \\
0 & 1 
\end{pmatrix}
\;\;\implies {\mathcal A}(z) = \cfrac{1}{\tau + \cfrac{1+k^2 \tau^2}{v-k^2 \tau + z}}\,.
$$
\end{enumerate}
\label{linearFractionalExercise}
\end{exercise}

\begin{exercise} 
In Exercise \ref{frischLloydExercise1}, we expressed the general solution of the Frisch--Lloyd model
for $E=-k^2 <0$, with $k>0$,
in terms of a product of random matrices where $A_j$ is given by Equation (\ref{negativeEnergyFrischLloydMatrix}).
Derive the formula
$$
{\mathcal A}_j^{-1} (z) = \frac{z \,\text{ch}(k \ell_j) - k \,\text{sh} (k \ell_j)}{\text{ch} (k \ell_j) - z \,\text{sh} (k \ell_j)/k}-v_j\,,
$$
and show that
$$
\frac{\partial}{\partial \ell_j} {\mathcal A}_j^{-1} (z) = (z^2-k^2) \frac{{\rm d} {\mathcal A_j}^{-1} (z)}{{\rm d} z}\,.
$$
\label{frischLloydExercise2}
\end{exercise}
\begin{center}
\linethickness{2mm}
\color{light-grey}{\line(1,0){360}}
\end{center}

In terms of the Riccati variable $Z$, we can therefore express
$$
\begin{pmatrix}
\psi'(x_{j+1}-) \\
\psi (x_{j+1})
\end{pmatrix}
= A_j 
\begin{pmatrix}
\psi'(x_{j}-) \\
\psi (x_{j})
\end{pmatrix}
$$
as
\begin{equation*}
Z(x_{j}-) = {\mathcal A}_j^{-1} \left ( Z (x_{j+1}-) \right ) \quad \text{for $j \ge 0$}\,.
\end{equation*}
For the impurity model, the general solution of the Riccati equation is therefore given by
\begin{equation}
Z(x_{n+1}-) = {\mathcal A}_n \circ \cdots \circ {\mathcal A}_1 \circ {\mathcal A}_0( Z(0-) )
\label{generalRiccatiSolution}
\end{equation}

In particular, if $L= x_{n+1}$, then
\begin{equation}
Z(0-) = {\mathcal A}_0^{-1} \circ {\mathcal A}_1^{-1} \circ \cdots \circ {\mathcal A}_n^{-1} \left ( z \right )\,.
\label{riccatiSolution}
\end{equation}
The right-hand side in  this last equation may be written as a finite continued fraction.

\begin{center}
\linethickness{2mm}
\color{light-grey}{\line(1,0){360}}
\end{center} 
\begin{exercise}
Set $L=x_{n+1}$. Show the following:
\begin{enumerate}[label=(\alph*)]
\item For Dyson's string,
$$
\frac{\chi_L'(0,\lambda)}{\chi_L(0,\lambda)} = \cfrac{1}{-\ell_0+\cfrac{1}{\lambda m_1 + \cfrac{1}{-\ell_1 +\ddots + \cfrac{1}{\lambda m_n + \cfrac{1}{-\ell_n+\cfrac{1}{z}}}}}}\,.
$$

\item For the Frisch--Lloyd model,
\begin{equation}
\frac{\chi_L'(0,\lambda)}{\chi_L(0,\lambda)}
= \cfrac{1}{-\tau_0+\cfrac{1+k^2 \tau_0^2}{k^2 \tau_0 -v_1 + \cfrac{1}{-\tau_1 +\ddots + \cfrac{1+k^2 \tau_{n-1}^2}{k^2 \tau_{n-1} -v_n + \cfrac{1}{-\tau_n+\cfrac{1}{z}}}}}}
\notag
\end{equation}
where $\tau_j := \tan (k \ell_j)/k$.
\end{enumerate}
\label{continuedFractionExercise}
\end{exercise}
\begin{center}
\linethickness{2mm}
\color{light-grey}{\line(1,0){360}}
\end{center}

\begin{remark}
It is clear from the foregoing discussion that 
$$
\frac{\chi_L'(0,\lambda)}{\chi_L (0,\lambda)}
$$
is {\em independent of $\alpha$}. Thus, although the Weyl coefficient itself does depend on the boundary 
condition at $x=0$, Exercise \ref{weylRiccatiExercise} shows that the dependence is trivial.
\label{alphaWeylRemark}
\end{remark}

\subsection{Classification in terms of limit-circle and limit-point types}
\label{classificationSubsection}
In order to understand what happens to the Weyl coefficient of the truncated problem 
as we take the limit $L \rightarrow \infty$,
it is helpful to allow the parameter $z$ to assume {\em complex} values. 

\begin{notation}
$$
{\mathbb C}_+ := \left \{ x + {\rm i} y :\, x \in {\mathbb R} \;\;\text{and}\;\; y > 0 \right \} \,.
$$
$$
\overline{{\mathbb C}_+} := \left \{ x + {\rm i} y :\, x \in {\mathbb R} \;\;\text{and}\;\; y \ge 0 \right \} \cup \{ \infty \}\,.
$$
\end{notation}

Although
the problem consisting of the equation (\ref{truncatedMasterEquation}) and the boundary conditions
(\ref{boundaryCondition}) is no longer self-adjoint when $z$ has a non-zero imaginary part, the
particular solutions $\psi(\cdot,\lambda)$, $\varphi(\cdot,\lambda)$ and $\chi_L(\cdot,\lambda)$ remain well-defined.
We set
\begin{equation}
w = w_L (\lambda) = -\frac{\varphi'(L-,\lambda)-z\,\varphi(L,\lambda)}{\psi'(L-,\lambda)-z\,\psi(L,\lambda)}
\label{definitionOfw}
\end{equation}
and view $w$ as a function of the complex parameter $z$. 

\begin{center}
\linethickness{2mm}
\color{light-grey}{\line(1,0){360}}
\end{center}
\begin{exercise}
Show that
\begin{equation*}
\text{Im} \,\lambda \,   \| \chi_L (\cdot,\lambda)  \|^2   = 
\text{Im} \, w - \text{Im} \,z\,\left | \chi_L(L,\lambda) \right |^2\,.
\end{equation*}
{\em Help:} Use Green's identity.
\label{integrabilityExercise}
\end{exercise}
\begin{center}
\linethickness{2mm}
\color{light-grey}{\line(1,0){360}}
\end{center}

This last equation says
\begin{multline}
\text{Im} \,\lambda \,\int_0^L \left | \varphi(x,\lambda) + w\, \psi(x,\lambda) \right |^2 \,{\rm d} M(x) -\text{Im}\,w \\
 = -\text{Im} \,z\, \left | \varphi(L,\lambda) + w\, \psi(L,\lambda) \right |^2\,.
\label{weylIdentity}
\end{multline}
In particular, 
$$
\text{Im} \,\lambda \,\int_0^L \left | \varphi(x,\lambda) + w\, \psi(x,\lambda) \right |^2 \,{\rm d} M(x) -\text{Im}\,w = 0 \;\;
\text{for $z \in {\mathbb R} \cup \{\infty\}$}\,.
$$
We claim that this is the equation of a {\em circle}, say $\partial D_L$, in the complex $w$-plane. To see this, we first rewrite
Equation (\ref{definitionOfw})  in the form
$$
z = \frac{\varphi'(L-,\lambda) + w \,\psi'(L-,\lambda)}{\varphi(L,\lambda) + w \,\psi(L,\lambda)}\,.
$$
Then the equation $\text{Im}\, z =0$ in the complex $z$-plane corresponds to the following equation
in the complex $w$-plane:
$$
\text{Im} \left [ \left ( \varphi' + w \,\psi' \right ) \left ( \overline{\varphi} + \overline{w} \,\overline{\psi} \right ) \right ] = 0
$$
where, for convenience, we momentarily omit 
to make explicit the dependence of $\varphi$ and $\psi$ on $L$ and $\lambda$. An equivalent form
of this equation is
$$
\left ( \varphi' + w \,\psi' \right ) \left ( \overline{\varphi} + \overline{w} \,\overline{\psi} \right ) 
= \left ( \overline{\varphi'} + \overline{w} \,\overline{\psi'} \right ) \left ( {\varphi} + {w} \,{\psi} \right ) \,.
$$
This, in turn, may be expressed as 
$$
\left | w - c_L \right |^2 = R_L^2
$$
where
\begin{equation}
c_L := \frac{\overline{\psi'(L-,\lambda)} \varphi (L,\lambda)-\varphi'(L-,\lambda) \overline{\psi(L,\lambda)}}{\psi'(L-,\lambda) \overline{\psi(L,\lambda)}-\overline{\psi'(L-,\lambda)} \psi(L,\lambda)}
\label{center}
\end{equation}
and
\begin{equation}
R_L := \left | \frac{\psi'(L-,\lambda) \varphi(L,\lambda)-\varphi'(L-,\lambda)\psi(L,\lambda)}{\psi'(L-,\lambda) \overline{\psi(L,\lambda)}-\overline{\psi'(L-,\lambda)} \psi(L,\lambda)} \right |\,.
\label{radius}
\end{equation}
Hence our claim.

\begin{center}
\linethickness{2mm}
\color{light-grey}{\line(1,0){360}}
\end{center}
\begin{exercise}
Show that
$$
\frac{1}{R_L} = 2 \,\left | \text{Im} \,\lambda \right | \int_0^L \left | \psi (x,\lambda) \right |^2 \,{\rm d} M(x)\,.
$$
{\em Help:} Use Green's identity to evaluate the numerator and the denominator on the right-hand side of Equation (\ref{radius}).
\label{centerExercise}
\end{exercise}
\begin{center}
\linethickness{2mm}
\color{light-grey}{\line(1,0){360}}
\end{center}

For definiteness, let us from now on assume that $\lambda \in {\mathbb C}_+$, and
denote by $D_L$ the open disk in the $w$-plane that has $\partial D_L$ as its boundary. We remark that
$w \in D_L$ if and only if
$$
\text{Im} \,\lambda \,\int_0^L \left | \varphi(x,\lambda) + w\, \psi(x,\lambda) \right |^2 \,{\rm d} M(x) -\text{Im}\,w < 0\,.
$$
Let us show that
$$
L' \ge L \implies D_{L'} \subseteq D_L\,.
$$
Indeed, if $L' \ge L$ and $w \in D_{L'}$ then
\begin{multline}
\notag
0> \text{Im} \,\lambda \,\int_0^{L'} \left | \varphi(x,\lambda) + w\, \psi(x,\lambda) \right |^2 \,{\rm d} M(x) -\text{Im}\,w \\
 \ge \text{Im} \,\lambda \,\int_0^{L} \left | \varphi(x,\lambda) + w\, \psi(x,\lambda) \right |^2 \,{\rm d} M(x) -\text{Im}\,w\,.
 \end{multline}
The geometrical implication is that $\{ D_L \}_{L > 0}$ is a {\em nested family} of disks, so that the
limits
$$
R := \lim_{L \rightarrow \infty} R_L \;\;\text{and}\;\; \partial D  := \lim_{L \rightarrow \infty} \partial D_L
$$
are well-defined.

Two possibilities therefore arise as $L \rightarrow \infty$:
\begin{enumerate}
\item $\partial D$ is a circle of radius $R>0$. In this case, 
we say that the point at infinity is of  {\em limit-circle} type. 
This implies among other things that  the
solution
$$
\varphi (\cdot,\lambda) + w \,\psi(\cdot,\lambda) 
$$
belongs to $L_M^2(0,\infty)$ for every $w \in D$. One can therefore assert that {\em every} solution belongs to $L_M^2(0,\infty)$.

\item $R = 0$ and $\partial D$ consists of a single point:
\begin{equation}
w(\lambda) := \lim_{L \rightarrow \infty} w_L(\lambda)\,.
\label{weylCoefficientLimit}
\end{equation}
In this case, we say that the point at infinity is of {\em limit-point} type.
Importantly, the limit (\ref{weylCoefficientLimit}) 
is then {\em independent of $z \in \overline{\mathbb C}$}. The function
\begin{equation}
\chi (\cdot,\lambda) := \varphi(\cdot,\lambda) + w(\lambda)\,\psi(\cdot,\lambda)
\label{weylSolutionLimit}
\end{equation}
is, up to a factor, the {\em only} solution of Equation (\ref{masterEquation}) that belongs to $L_M^2(0,\infty)$.
\end{enumerate}

Now that we have explained this classification,
let us return to the spectral problem for the infinite interval. For our particular impurity models, we make the following assumptions:
\begin{equation}
\tag{P}
\begin{cases}
\lim_{j \rightarrow \infty} x_j = \infty  & \\
 & \\
\{ m_j \}_{j \in {\mathbb N}} \subset {\mathbb R}_+\;\; \text{(Dyson)}\;\; \text{or} \;\;\{ v_j \}_{j \in {\mathbb N}} \subset {\mathbb R}_+\;\; \text{(Frisch--Lloyd)}   &
\end{cases}\,.
\label{limitPointHypothesis}
\end{equation}
Under this assumption, it is known that the point at infinity is of limit-point type when $\text{Im} \,\lambda > 0$;
see Kac \& Kre\u{\i}n (1974), \S 12, for Dyson's string, 
and Coddington \& Levinson (1955), Chapter 9, Theorem 2.4, for the Frisch--Lloyd model.
Whatever $z \in {\mathbb R} \cup \{ \infty\}$ one chooses to specify the boundary condition
at $x=L$ in the truncated problem, the large-$L$ limit will correspond
to imposing the integrability
condition
\begin{equation}
\psi \in L_M^2 (0,\infty)\,.
\label{interabilityCondition}
\end{equation}

In particular,
there is a limit measure $\sigma$ such that
\begin{equation}
\notag
\int_{\mathbb R} \eta (\lambda) \, {\rm d} \sigma_L (\lambda) \xrightarrow[L \rightarrow \infty]{} \int_{\mathbb R} \,\eta (\lambda) \,{\rm d} \sigma (\lambda) 
\end{equation}
for every smooth function $\eta$ with compact support.
The measure $\sigma$ is called the spectral measure
of the singular problem (\ref{masterEquation}); we denote its support by $\Sigma$.

The spectral measure can, as in the truncated case, be recovered from the Weyl coefficient:  Theorem \ref{coddingtonLevinsonTheorem}
remains valid in the limit as $L \rightarrow \infty$ and so we have
\begin{equation}
\sigma(\lambda) - \sigma(\lambda_0) = 
\frac{1}{\pi}  \int_{\lambda_0}^\lambda {\rm d} \zeta  \,\text{Im} \,w (\zeta+{\rm i} 0 )
\label{stieltjesFormula}
\end{equation}
at points of continuity $\lambda$ and $\lambda_0$ of $\sigma$.

\begin{center}
\linethickness{2mm}
\color{light-grey}{\line(1,0){360}}
\end{center}
\begin{exercise}
Consider the case $Q' = 0$ and $M' = 1$.
\begin{enumerate}[label=(\alph*)]
\item For $\lambda = - k^2$ with $k > 0$, show that
$$
w (\lambda) = \frac{\cos \alpha - k \sin \alpha}{k \cos \alpha + \sin \alpha}\,.
$$
\item Deduce that
$$
\sigma' (\lambda) = \frac{1}{\pi} \,\frac{\sqrt{\lambda}}{\sin^2 \alpha + \lambda \cos^2 \alpha}\,,\;\;\lambda >0\,.
$$
\end{enumerate}
\label{freeSpectralExercise}
\end{exercise}
\begin{center}
\linethickness{2mm}
\color{light-grey}{\line(1,0){360}}
\end{center}

As the exercise illustrates, unlike $\sigma_L$, the spectral measure $\sigma$ need not consist only of point masses. In general, one has the decomposition
\begin{equation}
\text{supp} \,\sigma =: \Sigma = \Sigma_{\text{p}} \cup \Sigma_{\text{a}} \cup \Sigma_{\text{s}}
\label{lebesgueDecomposition}
\end{equation}
where the sets on the right-hand side are, respectively, the {\em point}, the {\em absolutely continuous}, and the {\em singular continuous} spectrum. The point spectrum contains the eigenvalues, i.e. the values of $\lambda$ for which there
is a non-zero solution of Equation (\ref{masterEquation}) that is normalisable. The absolutely continuous spectrum consists of the values of $\lambda$ for which the derivative $\sigma'$ exists in the usual sense. The singular continuous spectrum is what is left after $\Sigma_{\text{p}}$ and $\Sigma_{\text{a}}$ have been removed.
It is generally difficult to prove that a measure is singularly continuous; an example of a measure which is conjectured to be such is given in \S \ref{furstenbergSection}, Example \ref{randomFibonacciExample}.

One further important consequence of the foregoing discussion is that, for the impurity models we have considered, the limit-point case arises whenever a certain continued fraction (see Exercise \ref{continuedFractionExercise}) converges as $n \rightarrow \infty$ to a limit independent of $z \in \overline{\mathbb C}$. 
This yields ``explicit'' formulae for the Weyl coefficient:
\begin{equation}
\frac{\chi_L'(0,\lambda)}{\chi_L(0,\lambda)}  = 
\cfrac{1}{-\ell_0+\cfrac{1}{\lambda m_1 + \cfrac{1}{-\ell_1 +\ddots}}} \;\; \text{(Dyson)}
\label{dysonWeylCoefficient}
\end{equation}
and
\begin{equation}
\frac{\chi_L'(0,\lambda)}{\chi_L(0,\lambda)}  = 
\cfrac{1}{-\tau_0 +\cfrac{1+k^2 \tau_0^2}{k^2 \tau_0-v_1 + \cfrac{1}{-\tau_1 +\ddots}}} \;\; \text{(Frisch--Lloyd)}
\label{frischLloydWeylCoefficient}
\end{equation}
where $\tau_j := \tan(k \ell_j)/k$.  

\begin{remark}
The continued fraction (\ref{dysonWeylCoefficient}) is essentially of the type studied by Stieltjes (1894; 1895) in connection with the solution of the moment problem for a measure supported on ${\mathbb R}_+$. 
\label{inverseRemark}
\end{remark}

\begin{remark}
\label{spectrumRemark}
The spectrum depends on
$M$ and $Q$, and so in the random case, it will depend on the particular realisation of those processes. 
Nevertheless, for the disordered models
that we consider here, it may be shown (see Pastur (1980)) that the spectrum is the same for almost every realisation. 
We refer to this common, non-random spectrum, as {\em the} spectrum, and use the letter
$\Sigma$ to denote it. 
\end{remark}

\begin{remark}
The Weyl coefficient $w$ is an analytic function of $\lambda$ in the upper half-plane. It may, however, have simple poles on the real line.
For instance, in Example \ref{freeSpectralExercise}, $w$ has a simple pole at $\lambda = -\tan^2 \alpha$ unless $\alpha=0$ or $\pi/2$.
\label{poleRemark}
\end{remark}

\subsection{Dyson's disordered strings of Type I}
\label{typeISubsection}
Let us now give an example of a random string for which the distribution of the Weyl coefficient may be found explicitly.

In his paper, Dyson discusses two kinds of disorder, which he calls ``Type I'' and ``Type II''. Strings of the latter type are obtained by choosing the $m_j$ and $\ell_j$ independently; such strings will be studied later in these notes. Strings of Type I are constructed as
follows: Let $c_0$, $c_1$, \ldots be positive random variables that are independent and identically distributed. 
Set
$$
\ell_0 := \frac{1}{c_0}\,, \; m_1 := \frac{c_0}{c_1}\,,\; \ell_1 := \frac{c_1}{c_0 \,c_2}\,,\;
m_2 := \frac{c_0 c_2}{c_1 c_3}\;\;  \text{etc.}
$$
With this definition, the $\ell_j$ and the $m_j$  are obviously
{\em dependent}.

\begin{center}
\linethickness{2mm}
\color{light-grey}{\line(1,0){360}}
\end{center}
\begin{exercise}
Set $\alpha = \pi/2$. Show that, for $\lambda \notin {\mathbb R}_+$, the Weyl coefficient of the Type I string is given by
$$
\frac{w(\lambda)}{\lambda} = Y \left (-1/\lambda \right )
$$
where
\begin{equation*}
Y(t) := \cfrac{c_0 t}{1+ \cfrac{c_1t}{1+ \cfrac{c_2 t}{1 + \ddots}}}\,.
\end{equation*}
\label{TypeIExercise1}
\end{exercise}
\begin{center}
\linethickness{2mm}
\color{light-grey}{\line(1,0){360}}
\end{center}
The method devised by Dyson for finding the integrated density of states of the random string
uses this continued fraction. He found the distribution of $Y(t)$ when $t>0$ and the $c_j$ have a gamma distribution with a certain choice of the parameters.
\begin{notation}
We say that the positive random variable $X$ has a $\text{\rm Gamma}(p,q)$ distribution if its probability density is of the form
$$
\frac{1}{q^p \Gamma (p)} x^{p-1} \exp \left ( -x/q \right )\,{\mathbf 1}_{(0,\infty)} (x)
$$ 
for some $p,\,q > 0$.
We say that the random variable $Y$ has a $\text{\rm Kummer}(p,q,r)$ distribution if its probability density is of the form
$$
\frac{1}{\Gamma(p) \Psi (p,1-q;r)} y^{p-1} (1+y)^{-p-q} {\rm e}^{-r y} {\mathbf 1}_{(0,\infty)} (y)
$$
for some $p > 0$, $q \in {\mathbb R}$ and $r > 0$, where $\Psi$ denotes the Kummer function. 
We use the $\sim$ symbol to indicate the law of a random variable, i. e.
$$
X \sim \text{Gamma}(p,q) \;\;\text{and}\;\;Y \sim \text{Kummer} (p,q,r)\,.
$$
\end{notation}

We state here Letac's generalisation of Dyson's result: 
\begin{multline}
\notag
X \sim \text{Gamma}(p,1/r) \;\;\text{and}\;\; Y \sim \text{Kummer} (p,q,r) \\
\implies  \cfrac{X}{1+Y} \sim \text{Kummer} (p+q,-q,r)
\end{multline}
where it is assumed that $X$ and $Y$ are independent.
A proof may be found in Letac (2009).

\begin{notation}
Let $X$ and $Y$ be two random variables. The notation
$$
X \overset{\text{law}}{=} Y
$$
means that $X$ and $Y$ have the same distribution.
\end{notation}

Letac's result therefore implies in particular that
\begin{equation}
\notag
X \sim \text{Gamma}(p,1/r) \;\;\text{and}\;\; Y \sim \text{Kummer} (p,0,r) 
\implies  Y \overset{\text{law}}{=} \cfrac{X}{1+Y} \,.
\end{equation}
Hence
\begin{equation}
c_j \sim \text{Gamma}(p,q) \implies \frac{w(\lambda)}{\lambda} \sim \text{Kummer} \left ( p,0,-\lambda/q \right )\,.
\label{dysonLetacFormula}
\end{equation}

\section{The complex Lyapunov exponent}
\label{dysonSchmidtSection}

For $\text{Im} \,\lambda > 0$, the particular solutions $\psi(\cdot,\lambda)$ and $\chi (\cdot,\lambda)$ form a basis for the
solution space of the differential equation (\ref{masterEquation}), and so every solution $\psi(x)$ is a linear combination of them:
$$
\psi(x) = a \,\psi (x,\lambda) + b\,\chi(x,\lambda)\,.
$$
In the limit point case, the only solutions  that
belong to $L_M^2 (0,\infty)$ are those for which $a =0$.
The other solutions do not decay sufficiently
quickly as $x \rightarrow \infty$ and, for them, we have
$$
\psi (x) \sim  a \,\psi(x,\lambda) \;\;\text{as $x \rightarrow \infty$}\,.
$$
In particular, if $| \psi(\cdot,\lambda) |$ grows exponentially, then $| \psi |$ will grow {\em at the same rate}.

\begin{definition}
The (real) {\em Lyapunov exponent} of the system (\ref{masterEquation}) is 
\begin{equation}
\gamma (\lambda) := \lim_{L \rightarrow \infty} {\mathbb E} \left ( \frac{\ln \left | \psi (L,\lambda) \right |}{L} \right )\,, \;\; \text{\rm Im} \,\lambda >0\,,
\label{lyapunovExponent}
\end{equation}
where the expectation is over $Q$ and $M$. 
\label{lyapunovExponentDefinition}
\end{definition}
In words, the Lyapunov exponent measures the expected rate of exponential growth of the particular solution $\psi (\cdot,\lambda)$,
and hence of every solution that is not proportional to $\chi(\cdot,\lambda)$.

\begin{remark}
Although the particular solution $\psi(\cdot,\lambda)$ defined by (\ref{fundamentalSolutions}) depends on the (real) parameter
$\alpha$ which specifies the boundary condition at $x=0$ for the spectral problem (\ref{masterEquation}), it is clear from the foregoing discussion
that the Lyapunov exponent itself is, for $\text{\rm Im}\,\lambda > 0$, independent of $\alpha$.
\label{alphaRemark}
\end{remark}

\begin{remark}
\label{selfAveragingRemark}
For our disordered model, it will turn out that the limit 
$$
 \lim_{L \rightarrow \infty} \frac{\ln \left | \psi (L,\lambda) \right |}{L}\,,
$$ 
whose expectation appears in the definition of the Lyapunov exponent, is the same for almost every realisation
of the disorder. Thus, we have here another example of a random variable that is, in the terminology of  \S \ref{motivationSubsection},  {\em self-averaging}. The Lyapunov exponent and the integrated density of states are self-averaging; {\em by contrast, the spectral measure is not.} This may be explained  heuristically by the fact that, as pointed out earlier, the spectral measure of the truncated problem
contains information about the norms of the eigenfunctions $\psi(\cdot,\lambda_j)$.
\end{remark}

\begin{remark}
\label{absolutelyContinuousRemark}
We emphasise that the Lyapunov exponent is {\em not defined in the spectrum}, where
the function $\psi(\cdot,\lambda)$ has its zeroes. Nevertheless, 
the limit $\gamma (\lambda + {\rm i} 0)$, $\lambda \in { \Sigma}$,
is meaningful, and contains valuable spectral information. 
We state without proof the following fact:  {\em Intervals where $\gamma(\lambda+\text{\rm i} 0)$ does
not vanish cannot belong to the absolutely continuous spectrum.} In particular,
it may be shown (Kotani 1982) that
$$
\forall \; \lambda \in \Sigma, \; \gamma(\lambda+ {\rm i} 0) > 0 \implies \Sigma_{ac} = \emptyset\,.
$$
Heuristically, if the absolutely continuous spectrum is empty, then one might expect the spectrum to be purely punctual; the eigenfunctions
must decay sufficiently rapidly to belong to $L_M^2 (0,\infty)$. 
More concretely, one may think of $\gamma(\lambda+{\rm i}0)$ as the rate of (exponential) decay of the eigenfunction corresponding
to $\lambda$, and of the reciprocal as a measure
of the spatial extent of the eigenfunction's support. In the literature on disordered systems, one often refers to $1/\gamma$
as the {\em localisation length}.
\end{remark}

It will be helpful to work with the following ``complex version'' of the Lyapunov exponent:
\begin{definition}
The function
\begin{equation}
\Omega (\lambda) := 
\lim_{L \rightarrow \infty} {\mathbb E} \left ( \frac{\ln  \psi (L,\lambda) }{L} \right )
\label{realLyapunov}
\end{equation}
is called the {\em complex Lyapunov exponent} of the system (\ref{masterEquation}).
\label{complexLyapunovDefinition}
\end{definition}
Obviously,
$$
\gamma = \text{Re} \,\Omega\,.
$$

\subsection{The imaginary part}
\label{imaginarySubsection}
Let us now discuss the significance of the imaginary part; for convenience, we set $\alpha = \pi/2$.
From the result of Exercise \ref{stringSolutionExercise}, 
we deduce that, for Dyson's model,
\begin{equation}
\psi(x_{n+1},\lambda) = c_n\,\left ( \lambda_{1} - \lambda \right ) \cdots
\left ( \lambda_{n} - \lambda \right )\,.
\label{stringSolution}
\end{equation}
The $\lambda_j$ are therefore the eigenvalues of Dyson's problem truncated to the interval $[0,L]$
with $L = x_{n+1}$ and $z=\infty$.

Let $\lambda > 0$, and suppose that it does not equal any of the $\lambda_{j}$. By virtue of Equation (\ref{stringSolution}), we then have
\begin{multline}
\notag
\ln \psi (x_{n+1},\lambda+{\rm i} 0) = \ln c_n + \sum_{\lambda_{j} < \lambda}  \left ( -{\rm i} \pi + \ln \left | \lambda_{j}-\lambda \right | \right ) 
+ \sum_{\lambda_{j} > \lambda} \ln   \left | \lambda_{j}-\lambda \right |  \\
=  \ln \left | \psi (x_{n+1},\lambda) \right | - {\rm i} \pi \sum_{\lambda_{j} < \lambda}  1 \,.
\end{multline}
We deduce
$$
\ln \psi (x_{n+1},\lambda+{\rm i} 0) = -{\rm i} \pi \,x_{n+1} N_{x_{n+1}} (\lambda) + \ln \left | \psi (x_{n+1},\lambda) \right |
$$
where $N_L(\lambda)$ is the counting measure defined by Equation (\ref{dysonCountingMeasure}).
Divide both sides  by $x_{n+1}$, and take the limit as $n \rightarrow \infty$. By definition of the complex
Lyapunov exponent and of the integrated density of states, we obtain
\begin{equation}
\text{Im} \,\Omega (\lambda + {\rm i} 0)  =   - \pi N(\lambda) \,.
\label{integratedDensityOfStatesFormula}
\end{equation}
For $\lambda$ ``just above'' the spectrum, the imaginary part of the complex Lyapunov exponent
therefore yields the integrated density of states that Dyson set out to calculate.

Formula (\ref{integratedDensityOfStatesFormula}) remains valid for the Frisch--Lloyd model, but the proof
is more complicated. A justification will be given later in \S \ref{furtherSection}.

\begin{remark}
As will become manifest in \S \ref{furtherSection}, the integrated density of states per unit length
$N(\lambda)$ does not depend on the boundary condition at $x=0$--- and hence on the parameter $\alpha$.
This fact, together with our earlier comments, makes it clear that the spectral information contained in the complex Lyapunov exponent
$\Omega$ is much coarser than that contained in the Weyl coefficient $w$.
\label{densityOfStatesRemark}
\end{remark}

\subsection{Some deterministic examples}
\label{deterministicSubsection}

$\Omega$ has the advantage of being {\em analytic} outside the spectrum, except at points where the Weyl coefficient
happens to have poles; this important observation, which goes back to Dyson himself (Dyson 1953) and has been exploited by others (see for instance Luck (1992)), can greatly facilitate its calculation.

\begin{example}
We illustrate this point by revisiting the simple but instructive case
where $Q' = 0$ and $M' =1$. In this (deterministic) case, $\Sigma \subset {\mathbb R}_+$,
and so the complex Lyapunov exponent is analytic along the negative half-line. Let us first compute $\Omega$ there: 
Setting $\lambda = - k^2 <0$, with $k > 0$, in Equation (\ref{masterEquation}) gives
$$
- \psi'' = -k^2 \psi\,.
$$
Therefore
$$
\psi(x,-k^2) = \cos \alpha \, \text{ch}(kx) + \frac{\sin \alpha}{k} \,\text{sh} (kx)
$$
and so, for $k \ne -\tan \alpha$,
$$
 \Omega(-k^2) = \lim_{L \rightarrow \infty} 
\frac{\ln \psi(L,-k^2)}{L} = k\,.
$$
By using
$$
k = \sqrt{-\lambda}
$$
we can continue this formula for the complex Lyapunov exponent to other values of the spectral parameter.
In particular, for $\lambda > 0$, we find
$$
\Omega( \lambda + {\rm i} 0) = -{\rm i} \sqrt{\lambda}\,.
$$
Hence, with the help of Formula (\ref{integratedDensityOfStatesFormula}), we easily recover the result obtained previously in Exercise \ref{weylRiccatiExercise} (c). Furthermore
$$
\gamma (\lambda + {\rm i} 0) = \text{Re} \,\Omega (\lambda+{\rm i} 0) = 0\,.
$$
\label{analyticContinuationExample}
\end{example}

The fact that, in this example, the real Lyapunov exponent vanishes along the spectrum is consistent with the fact that the spectral measure is absolutely continuous. Another less trivial case where this result can be verified is that of the Kronig-Penney model discussed in \S \ref {frischLloydSubsection}. The band spectrum  $\Sigma$  is characterized by 
$$
\Sigma=\{\lambda \in\mathbb R :\, \gamma(\lambda+{\rm i} 0)=0\}=\{k^2 \in \mathbb R :\, |\cos (k \ell) +  \sin (k \ell) \,v/(2 k)| < 1\}\,.
$$
We shall see that, by contrast, in the {\em disordered} case, $\gamma(\lambda + {\rm i} 0)$ is strictly positive
everywhere in the spectrum; this is the mathematical manifestation, in the one-dimensional case, of the phenomenon known as {\em Anderson localisation}.

\begin{example}
Consider a deterministic homogeneous string such that all the masses equal $m$ and the spacing between two successive impurities
equals $\ell$. Let us calculate $\Omega$ explicitly for $\lambda < 0$. 

In this case, the general solution $\psi$ of Equation (\ref{masterEquation})
is a continuous function of $x$ that is piecewise linear with respect to the partition
$$
0 = x_0 < x_1 < \ldots\;\;\text{where}\; x_j = j \ell\,.
$$
The values taken at the $x_j$ satisfy the recurrence relation
$$
\psi(x_{j+1}) - 2 \psi(x_j) + \psi(x_{j-1}) = - \lambda m \ell \,\psi (x_j)\,,\;\; j =1,\,2,\, \ldots
$$
Hence
$$
\psi(x_j) = a \,\text{ch} ( \Omega x_j ) + b \,\text{sh} (\Omega x_j )
$$
where 
$$
{\rm e}^{\pm \Omega \ell} = 1 - \frac{\lambda m \ell}{2} \pm \sqrt{\left ( 1-\frac{\lambda m \ell}{2} \right )^2-1}\,.
$$
From this, we readily deduce exact expressions for the particular solutions $\psi(x,\lambda)$ and for $\chi(x,\lambda)$:
For instance, in the particular case of a Dirichlet boundary condition at $x=0$, i.e. $\alpha = \pi/2$, we have
$$
\chi(x,\lambda) = \frac{x_j-x}{\ell} {\rm e}^{-\Omega x_{j-1}} + \frac{x-x_{j-1}}{\ell} {\rm e}^{-\Omega x_j} \;\;\text{for} \; x_{j-1} \le x \le x_j
$$
and
$$
\psi(x,\lambda) = \frac{\ell}{\text{sh}(\Omega \ell)} \left [ \frac{x_j-x}{\ell} \text{sh} ( \Omega x_{j-1} ) + 
 \frac{x-x_{j-1}}{\ell} \text{sh} ( \Omega x_{j} ) \right ] \;\;\text{for} \; x_{j-1} \le x \le x_j\,.
$$
Therefore
$$
 \lim_{x \rightarrow \infty} \frac{\ln \psi (x,\lambda)}{x} = \Omega (\lambda) =
 \frac{1}{\ell} \ln \left [1 - \frac{\lambda m \ell}{2} + \sqrt{\left ( 1-\frac{\lambda m \ell}{2} \right )^2-1}  \right ]
$$
and
$$
w(\lambda) = \chi'(0,\lambda) = - \frac{1-{\rm e}^{-\Omega \ell}}{\ell}\,.
$$
\label{homogeneousStringExample}
\end{example}

\begin{center}
\linethickness{2mm}
\color{light-grey}{\line(1,0){360}}
\end{center}
\begin{exercise}
Consider the homogeneous string with $\alpha =\pi/2$.
\begin{enumerate}[label=(\alph*)]
\item Show that its spectral measure is supported on the interval
$0 < \lambda < 4/(m \ell)$, where it is given by
$$
\sigma'(\lambda) = \frac{1}{\pi \ell} \sqrt{1- \left ( 1-\lambda m \ell/2\right )^2}\,.
$$
\item Show that, for $0 < \lambda < 4/(m \ell)$,
$$
N(\lambda) =\frac{1}{\pi \ell} \arccos \left ( 1-\frac{\lambda m \ell}{2}  \right )\,.
$$
\end{enumerate}
\label{homogeneousStringExercise}
\end{exercise}
\begin{center}
\linethickness{2mm}
\color{light-grey}{\line(1,0){360}}
\end{center}

\subsection{A disordered example: Kotani's Type II string}
\label{randomStringSubsection}
We remark that, in
Example \ref{analyticContinuationExample}, $\Omega$ is the negative of the Weyl coefficient corresponding to a Dirichlet condition at $x=0$.
Example \ref{homogeneousStringExample} shows that
this relationship between the complex Lyapunov exponent and the Weyl coefficient does not hold in general.
Nevertheless, we proceed to give an example which suggests that a connection between these two objects 
does exist if the system is disordered.

Let 
$$
Q' = 0 \;\;\text{and}\;\;M'  =  \sum_{j=1}^\infty m_j \delta (x-x_j)\,.
$$ 
Kotani (1976), in his Example 2, \S 5, considered the case of a random string of Type II
where the $m_j$ and the $\ell_j$ are two sequences
of independent random variables with
\begin{equation}
m_j \sim \text{Gamma}(1,m) \;\;\text{and}\;\; \ell_j \sim \text{Gamma}(1,\ell)\,.
\label{kotaniExample}
\end{equation}
By a completely rigorous calculation, he found
\begin{equation}
N(\lambda) =  \frac{m \lambda/\pi^2}{J_{1}^2 \left ( \frac{2}{\sqrt{m \lambda \ell}} \right ) + Y_1^2\left ( \frac{2}{\sqrt{m \lambda \ell}}\right ) }\,.
\label{kotaniFormula}
\end{equation}
Our purpose in what follows is to show that the right-hand side is also the expected value of the spectral density $\sigma'$
for the particular choice $\alpha = \pi/2$.

As mentioned already, it is useful to work outside the spectrum. We 
let $\lambda<0$ and $\alpha = \pi/2$;
the continued fraction (\ref{dysonWeylCoefficient}) for $-w(\lambda)$ is then clearly a {\em positive} random variable.

\begin{notation}
We say that the positive random variable $X$ has a $\text{GIG}(p,q,r)$ (generalised inverse Gaussian) distribution if its probability
density function is of the form
$$
\frac{(q/r)^{p/2}}{2 K_p ( \sqrt{qr} )} x^{p-1} \exp \left [ -\frac{1}{2} \left ( q x + r/x\right ) \right ]\,{\mathbf 1}_{(0,\infty)}(x)
$$ 
for some non-negative numbers $p$, $q$ and $r$.
\end{notation}

Letac \& Seshadri (1983) showed the following: Let $a$, $b$ and $p$ be positive numbers, and let
$$
\Gamma_1 \sim \text{Gamma}(p,2/b)\,,\;\; \Gamma_2 \sim \text{Gamma}(p,2/a)\;\;\text{and} \;\; X \sim \text{GIG}(-p,a,b)
$$
be three mutually independent random variables. Then
\begin{equation}
X \overset{\text{law}}{=} \cfrac{1}{\Gamma_1 + \cfrac{1}{\Gamma_2 + X}}\,.
\label{letacSeshadri}
\end{equation}

When we apply this result to the random string of Type II (\ref{kotaniExample}), we find
$$
-w(\lambda) \sim \text{GIG} \left ( -p,a, b\right )\;\;\text{with}\;\; p =1,\, a = \frac{-2}{m \lambda}\;\text{and}\;
b= \frac{2}{\ell} \,.
$$

\begin{center}
\linethickness{2mm}
\color{light-grey}{\line(1,0){360}}
\end{center}
\begin{exercise}
Show that
\begin{equation}
\notag
{\mathbb E} \left [ -w(\lambda) \right ] = \left ( -\frac{m \lambda}{\ell} \right )^{\frac{1}{2}} \,\frac{K_0 \left ( \frac{2}{\sqrt{-m \lambda \ell}} \right )}{K_1 \left ( \frac{2}{\sqrt{-m \lambda \ell}} \right )} \quad \text{for $\lambda < 0$}\,.
\end{equation}
\label{firstLetacExercise}
\end{exercise}
\begin{center}
\linethickness{2mm}
\color{light-grey}{\line(1,0){360}}
\end{center}

Next, we use analytic continuation in $\lambda$: For $-\pi/2 \le \arg z \le \pi$,
$$
K_{\nu}(z) = -\frac{1}{2} \pi {\rm i} {\rm e}^{-{\rm i} \pi \nu/2} H_{\nu}^{(2)} \left ( z {\rm e}^{-{\rm i} \pi/2} \right )\,.
$$
Therefore
\begin{equation}
\notag
{\mathbb E} \left [ -w(\lambda+{\rm i} 0) \right ] = 
\left ( \frac{m \lambda}{\ell} \right )^{\frac{1}{2}} \,\frac{H_0^{(2)} \left ( \frac{2}{\sqrt{m \lambda \ell}} \right )}{H_1^{(2)} \left ( \frac{2}{\sqrt{m \lambda \ell}} \right )}
\quad \text{for $\lambda > 0$}\,.
\end{equation}
Now,
$$
2 \,\text{Im} \, \frac{H_0^{(2)}(z)}{H_1^{(2)}(z)} = 2 \,\text{Im} \, \frac{H_0^{(2)}(z) H_1^{(1)}(z)}{\left | H_1^{(1)}(z) \right |^2}
= \text{Im} \frac{H_0^{(2)}(z) H_1^{(1)}(z)-H_0^{(1)}(z) H_1^{(2)} (z)}{\left | H_1^{(1)}(z) \right |^2}\,.
$$
Equation 10.5.5, of NIST's Digital Library of Mathematical Functions, says that
$$
H_0^{(2)}(z) H_1^{(1)}(z)-H_0^{(1)}(z) H_1^{(2)} (z) = - \frac{4 {\rm i}}{\pi z}\,.
$$
Hence, after some re-arrangement, we obtain, for $\lambda > 0$,
$$
\frac{1}{\pi} {\mathbb E} \left [ w(\lambda+{\rm i} 0) \right ] = \frac{m \lambda}{\pi^2} \frac{1}{\left | H_1^{(1)}(z)\right |^2}\,.
$$
The right-hand side is the same as in Kotani's formula (\ref{kotaniFormula}). Hence, for $\lambda > 0$,
$$
{\mathbb E} \left [ w(\lambda+{\rm i} 0) \right ] = -\text{Im}\,\Omega( \lambda + {\rm i} 0)\;\;\text{for $\lambda > 0$}\,.
$$
Theorem \ref{coddingtonLevinsonTheorem} then implies
\begin{equation}
{\mathbb E} \left [ \sigma'(\lambda) \right ] = N(\lambda)\,.
\label{meanDensityFormula}
\end{equation}
We shall, in \S \ref{relationshipSubsection}, show that this relationship holds more generally provided that
certain conditions are satisfied.

\begin{center}
\linethickness{2mm}
\color{light-grey}{\line(1,0){360}}
\end{center}
\begin{exercise}
Show that, for Kotani's string,
$$
N (\lambda) \sim \left ( \frac{m}{\ell} \right )^{\frac{1}{2}} \frac{\sqrt{\lambda}}{\pi}
\;\;\text{as $\lambda \rightarrow 0+$}\,.
$$
This expresses the fact that, for a large wavelength, the density of states approaches the density of states corresponding to 
a homogeneous string (for which $M' > 0$ is independent of $x$).
\label{secondLetacExercise}
\end{exercise}
\begin{center}
\linethickness{2mm}
\color{light-grey}{\line(1,0){360}}
\end{center}

\subsection{Calculation of the Lyapunov exponent}
\label{calculationSubsection}

Just as was the case for the Weyl coefficient, the calculation of the complex Lyapunov exponent makes use of the  Riccati equation (\ref{riccatiEquation}) introduced earlier.
This time, however, we are interested in the particular solution
\begin{equation}
Z (x, \lambda) := \frac{\psi'(x,\lambda)}{\psi (x,\lambda)}
\label{riccatiVariable}
\end{equation}
which we shall refer to as the {\em Riccati process}.

\begin{quote}
{\bf Suppose that the ${\mathcal A}_j$, $j =1,2,\ldots$, are independent and with the same distribution $\mu$}. 
\end{quote}

By applying Formula (\ref{generalRiccatiSolution}) for the general solution of the Riccati equation,
we see that the random variables $\{ Z(x_j-,\lambda) \}_{j \in {\mathbb N}}$ form a
Markov chain:
\begin{equation}
Z(x_{j+1}-,\lambda) = {\mathcal A}_j \left ( Z(x_j-,\lambda) \right ) \;\; \text{for $j=1,2,\ldots$}\,.
\label{forwardIteration}
\end{equation}
This Markov chain has a stationary distribution; we denote the stationary probability measure by $\nu$.
When $\lambda$ is {\em real}, the support of $\nu$ is contained in ${\mathbb R}$; more generally, it is
contained in the complex plane. 

\begin{center}
\linethickness{2mm}
\color{light-grey}{\line(1,0){360}}
\end{center}
\begin{exercise}
Show that, if $\nu$ has a density supported on the real line, say 
$$
{\rm d} \nu(z) = f(z)\,{\rm d} z\,,
$$
then
\begin{equation}
f(z) = {\mathbb E} \left ( \left [ f \circ {\mathcal A}^{-1} \right ] (z) \frac{{\rm d} {\mathcal A}^{-1}}{{\rm d} z} (z) \right )
\label{letacEquation}
\end{equation}
where the expectation is over the $\mu$-distributed matrix random variable $A$.
\label{dysonSchmidtEquation}
\end{exercise}
\begin{center}
\linethickness{2mm}
\color{light-grey}{\line(1,0){360}}
\end{center}

This equation for the stationary density $f(z)$ of the Riccati process is known in the physics literature as
the {\em Dyson--Schmidt equation}.
There is no systematic method for solving it. Nevertheless, there are cases where the solution has been obtained explicitly.
It will be instructive in what follows to consider also its integrated version:
\begin{equation}
{\mathbb E} \left ( \int_z^{{\mathcal A}^{-1}(z)} {\rm d} t \, f(t) \right ) = J\,.
\label{integratedDysonSchmidtEquation}
\end{equation}
We shall see in due course that the constant of integration $J$ has a physical interpretation.

Returning to the calculation of the complex Lyapunov exponent, we have
\begin{equation}
\notag
\frac{1}{L} \ln \frac{\psi (L,\lambda)}{\psi (a,\lambda)} =
\frac{1}{L} \int_a^L Z(x,\lambda)\,{\rm d} x
\end{equation}
where $a>0$ is an arbitrary point.

\begin{quote}
{\bf We make the hypothesis that
the Riccati process behaves ergodically.}
\end{quote} 

Concretely, this means that we assume the following:
\begin{equation}
\Omega (\lambda) = \lim_{L \rightarrow \infty} \frac{1}{L} \int_a^L Z(x,\lambda)\,{\rm d} x 
= \int_{\text{supp} \,\nu} z \,{\rm d} \nu (z) \;\;\text{almost surely.}
\label{frischLloydFormula}
\end{equation}

\subsection{The relationship between $\Omega(\lambda)$ and $w(\lambda)$}
\label{relationshipSubsection}

In order to clarify the relationship between the complex Lyapunov exponent and the Weyl coefficient, we introduce, with Letac (1986), 
the so-called {\em backward
iterates}  $\{Z_{n+1}\}$ associated with the (forward) Markov chain $\{ Z(x_n-,\lambda) \}_{n \in {\mathbb N}}$:
\begin{equation}
Z_{n+1} := {\mathcal A}_1 \circ {\mathcal A}_{2} \circ \cdots \circ {\mathcal A}_n ( Z (x_1-,\lambda) )\,.
\label{definitionOfZn}
\end{equation}
Since, by assumption, the $A_j$ are independent and $\mu$-distributed, 
Formula (\ref{generalRiccatiSolution}) implies
\begin{equation}
Z_{n+1} \overset{\text{law}}{=} {\mathcal A}_n \circ {\mathcal A}_{n-1} \circ \cdots \circ {\mathcal A}_1 ( Z (x_1-,\lambda) )
= Z(x_{n+1}-,\lambda)\,.
\label{equalityInLaw}
\end{equation}
We have for these backward iterates the following counterpart of Exercise \ref{continuedFractionExercise}:

\begin{center}
\linethickness{2mm}
\color{light-grey}{\line(1,0){360}}
\end{center}
\begin{exercise}
Show the following:
\begin{enumerate}[label=(\alph*)]
\item For Dyson's string,
$$
Z_{n+1} = \cfrac{1}{\ell_1 + \cfrac{1}{-\lambda m_1 + \ddots + \cfrac{1}{\ell_n + \cfrac{1}{-\lambda m_n + Z(x_1-,\lambda)}}}}\,.
$$

\item For the Frisch--Lloyd model,
$$
Z_{n+1} = \cfrac{1}{\tau_1 + \cfrac{1+k^2 \tau_1^2}{v_1-k^2 \tau_1 + \ddots + \cfrac{1}{\tau_n + \cfrac{1+k^2\tau_n^2}{v_n -k^2 \tau_n + Z(x_1-,\lambda)}}}}
$$
where $\tau_j := \tan (k \ell_j)/k$.
\end{enumerate}
\label{forwardIterationExercise}
\end{exercise}
\begin{center}
\linethickness{2mm}
\color{light-grey}{\line(1,0){360}}
\end{center}

For every fixed $n$, the backward and forward iterates have the same law, but their
large-$n$ behaviours are very different; see Figure \ref{forwardBackwardFigure} for an illustration.
The $Z(x_n-,\lambda)$ behave ergodically. By contrast,
the $Z_n$ converge almost surely
to a random limit:
\begin{equation}
Z_\infty := \lim_{n \rightarrow \infty} {\mathcal A}_1 \circ {\mathcal A}_{2} \circ \cdots \circ {\mathcal A}_n ( Z(x_1-,\lambda) )\,.
\label{limitContinuedFraction}
\end{equation}
This limit is an infinite continued fraction independent of $Z(x_1-,\lambda)$:
\begin{equation}
Z_\infty = \cfrac{1}{\ell_1 + \cfrac{1}{-\lambda m_1 + \cfrac{1}{\ell_2 + \cfrac{1}{-\lambda m_2 + \ddots}}}} \quad \text{(Dyson)}
\label{dysonContinuedFraction}
\end{equation}
and
\begin{equation}
Z_\infty = \cfrac{1}{\tau_1 + \cfrac{1+k^2 \tau_1^2}{v_1 - k^2 \tau_1 + \cfrac{1}{\tau_2 + \cfrac{1+k^2 \tau_2^2}{v_2 -k^2 \tau_2 + \ddots}}}}  \quad \text{(Frisch--Lloyd)}\,.
\label{frischLloydContinuedFraction}
\end{equation}

Now, by Equation (\ref{equalityInLaw}), the law of $Z_\infty$ is 
the stationary distribution $\nu$ of the Markov chain.
Hence Formula (\ref{frischLloydFormula}) may be expressed alternatively as
\begin{equation}
\Omega = {\mathbb E} \left ( Z_\infty \right )\,.
\label{alternativeFrischLloydFormula}
\end{equation}
Since the $\ell_j$ are independent and identically distributed, as are the $m_j$ and the $v_j$, we deduce upon
comparison with the continued fractions
in Equations (\ref{dysonWeylCoefficient}) and (\ref{frischLloydWeylCoefficient}) that
\begin{equation}
Z_\infty \overset{\text{law}}{=} - \frac{\chi_L'(0,\lambda)}{\chi_L (0,\lambda)}\,.
\label{secondEqualityInLaw}
\end{equation}
For both of these impurity models, we therefore have
\begin{equation}
\Omega (\lambda) = - {\mathbb E} \left [ \frac{\chi_L'(0,\lambda)}{\chi_L (0,\lambda)} \right ]\,.
\label{lyapunovInTermsOfWeyl}
\end{equation}

This takes a particularly simple form when $\alpha = \pi/2$; we may then use Theorem 1 and take the expectation to deduce
$$
{\mathbb E} \left ( \sigma' \right ) = -\frac{1}{\pi} \text{Im}\,\Omega (\lambda+{\rm i} 0)\,.
$$
For Dyson's string, the right-hand side is the integrated density of states, and so Equation (\ref{meanDensityFormula}), obtained in the context of a specific random string, holds more generally.

\begin{remark}
We emphasise that the equality in law (\ref{secondEqualityInLaw}) is a nontrivial fact. 
For our particular
impurity models, it relies on the ``i.i.d.'' nature of the coefficients that appear in the continued fraction expansions,
and also
on the hypothesis that the Riccati process enjoys the ergodic property (\ref{frischLloydFormula}).
\label{weylRemark}
\end{remark}

\subsection{Application to the Frisch--Lloyd model}
\label{nieuwenhuizenSubsection}
For the Frisch--Lloyd model, we do not have ``ready-made'' formulae for the distribution of the random continued fractions
(\ref{frischLloydWeylCoefficient}) and (\ref{frischLloydContinuedFraction}). We must therefore work with the Riccati process, taking as our starting point the Dyson--Schmidt equation (\ref{letacEquation})
for the unknown probability density $f$. If we assume that the coupling constants $v_j$ are positive then the spectrum is contained in ${\mathbb R}_+$, and
it will again be convenient to work on the negative half-line $\lambda = -k^2<0$, where the Lyapunov exponent is analytic. 

\begin{notation}
For the sake of economy, we shall in what follows use the same symbols $\ell_j$ and $v_j$
to denote random variables 
or variables of integration.
\end{notation}

In view of the result obtained in Exercise \ref{frischLloydExercise2}, we can write the Dyson--Schmidt equation
in the form
\begin{equation}
(z^2-k^2) \,f(z) = \int_{\text{supp} \,\mu} {\rm d} \mu \, f \left ( {\mathcal A}_j ^{-1} (z) \right ) \frac{\partial {\mathcal A}_j^{-1} (z)}{\partial \ell_j}\,.
\label{dysonSchmidtForFrischLloyd}
\end{equation}

We now consider the special 
case where the $\ell_j$ are independent with a common exponential distribution with mean $\ell$, i.e.
for every measurable subset $S \subseteq {\mathbb R}_+$,
\begin{equation}
\notag
{\mathbb P} \left ( \ell_j \in S \right ) = \int_S \frac{1}{\ell} {\rm e}^{- x/\ell} \,{\rm d} x\,.
\end{equation}
$1/\ell$ is therefore the mean density of the impurities.
We suppose also that the $v_j$ are mutually independent random variables, independent also of the $\ell_j$, with a common distribution
whose density we shall denote by $\rho$.

\begin{center}
\linethickness{2mm}
\color{light-grey}{\line(1,0){360}}
\end{center}
\begin{exercise}
Set
$$
\Phi( \ell_j,v_j,z ) := \int_{z-v_j}^{{\mathcal A}_j^{-1} (z)} {\rm d} t f(t)\,.
$$
Show the following:
\begin{enumerate}[label=(\alph*)]
\item 
$$
(z^2-k^2) f(z) = \int_{\text{supp} \rho} {\rm d} v_j \,\rho(v_j) \int_0^\infty {\rm d} \ell_j \,\frac{1}{\ell}\,{\rm e}^{-\ell_j/\ell} \frac{\partial \Phi}{\partial \ell_j}\,.
$$

\item 
$$
(z^2-k^2) f(z) = \frac{1}{\ell} \,{\mathbb E} \left [  \Phi(\ell_j,v_j,z) \right ]\,.
$$

\item For every $z > 0$,
\begin{equation}
(z^2-k^2) f(z) + \frac{1}{\ell} \,\int_{\text{supp} \rho} {\rm d} v_j \,\rho(v_j)\, \int_{z}^{z-v_j} {\rm d} t\,f(t) =  \frac{J}{\ell}\,.
\label{frischLloydDensityEquation}
\end{equation}

\item Deduce that, if $Z_\infty$ has a mean, then $J=0$.
\end{enumerate}
\label{solvableFrischLloydExercise2}
\end{exercise}
\begin{center}
\linethickness{2mm}
\color{light-grey}{\line(1,0){360}}
\end{center}

The upshot of the exercise is that we have reduced the Dyson--Schmidt equation
to an equation that is {\em linear}. The presence of an integral term makes it difficult to solve directly, but
Frisch \& Lloyd (1960) observed that the problem simplifies if one works with the {\em Laplace transform}
\begin{equation}
F(p) := \int_{0}^\infty f(z) \,{\rm e}^{- p z}\,{\rm d} z\,.
\label{laplaceTransform}
\end{equation}

\begin{center}
\linethickness{2mm}
\color{light-grey}{\line(1,0){360}}
\end{center}
\begin{exercise}
Show that, for $\lambda=-k^2 <0$,
\begin{equation}
F'' -k^2 F+ \frac{1}{\ell} \,{\mathbb E} \left ( \frac{ {\rm e}^{- p v_j} -1}{p} \right ) \,F = 0\,.
\label{laplaceEquation}
\end{equation}
\label{laplaceExercise}
\end{exercise}
\begin{center}
\linethickness{2mm}
\color{light-grey}{\line(1,0){360}}
\end{center}

In terms of $F$, we have
$$
\Omega(-k^2) = -F'(0+)
$$
and so the problem reduces
to finding a solution of a {\em linear homogeneous differential equation}
that is positive and decays to zero as $p \rightarrow \infty$.

\begin{example}
Let the $v_j$ be independent and exponentially-distributed with mean $v>0$, i.e.
$$
\rho (x) = \frac{1}{v}\, {\rm e}^{-x/v} {\mathbf 1}_{(0,\infty)} (x) \,.
$$
In this case, Equation (\ref{laplaceExercise}) becomes
$$
F'' - k^2 F -  \frac{1/\ell}{p+1/v} F = 0\,.
$$
The general solution may be expressed in terms of the Whittaker functions; see NIST's Digital Library of Mathematical Functions, \S 13.14. The only decaying
solutions are those proportional to
$$
W_{\frac{-1}{2k\ell},\frac{1}{2}} \left ( 2 k (p+1/v) \right )\,.
$$
Hence
$$
\Omega (\lambda) = -2 \sqrt{-\lambda} \frac{W_{\frac{-1}{2\ell \sqrt{-\lambda}},\frac{1}{2}}' \left ( 2/v \,\sqrt{-\lambda} \right )}{W_{\frac{-1}{2\ell \sqrt{-\lambda}},\frac{1}{2}} \left ( 2/v \,\sqrt{-\lambda}\right )} \quad \text{for $\lambda < 0$}\,.
$$
Analytic continuation then yields 
\begin{equation}
\Omega (\lambda+ {\rm i} 0) = 2 {\rm i} \sqrt{\lambda} \, \frac{W_{\frac{-{\rm i}}{2 \ell \sqrt{\lambda}},\frac{1}{2}} ' \left (  - 2 {\rm i} \sqrt{\lambda}/v \right )}{W_{\frac{-{\rm i}}{2 \ell \sqrt{\lambda}},\frac{1}{2}}\left (  -2 {\rm i} \sqrt{\lambda}/v \right )}
\quad \text{for $\lambda > 0$}\,.
\label{characteristicFunctionForDeltaScatterer}
\end{equation}
This formula for the characteristic function was discovered by Nieuwenhuizen (1983).
\label{nieuwenhuizenExample}
\end{example}

\section{Further remarks on the Frisch--Lloyd model}
\label{furtherSection}

For Dyson's string, we were able to show that the imaginary part of the complex
Lyapunov exponent yields the integrated density of states, i.e.
$$
- \frac{1}{\pi} \,\text{Im} \, \Omega (\lambda+{\rm i} 0) = N (\lambda) := \lim_{L \rightarrow \infty} N_L (\lambda)
$$
where $N_L$ is the counting measure
$$
N_L (\lambda) := \frac{\# \left \{ n \in {\mathbb N}:\, \lambda_n < \lambda \right \}}{L}\,.
$$
For the Frisch--Lloyd model, the function $\psi(L,\lambda)$ is {\em not} a polynomial
in $\lambda$, and the elementary approach used for Dyson's string is inapplicable. We shall instead proceed
in three steps: 
\begin{enumerate}
\item First, in \S \ref{pruferSubsection}, we point out the close relationship between the zeroes of 
$x \mapsto \psi(x,\lambda)$ and those of $\lambda \mapsto \psi(x,\lambda)$. 
\item Secondly, in \S \ref{frischLloydAnalysisSubsection}, we relate the distribution of these zeroes 
to the tail behaviour of the stationary probability density of the Riccati variable.
\item Finally, the connection with the complex Lyapunov exponent is made in \S \ref{riceSubsection}.
\end{enumerate}

For the sake of brevity, we restrict our attention to the Frisch--Lloyd model, i.e. $Q'=V$ and $M'=1$
and, for the truncated problem, will consider only the case where the boundary condition at $x=L$
is Dirichlet's, i.e. $z = \infty$.

\subsection{Pr\"ufer variables: the phase formalism}
\label{pruferSubsection}

\begin{theorem}[Sturm]
Let 
$$
-\infty < \lambda_1 < \lambda_2 < \cdots 
$$
denote the eigenvalues of the spectral problem for (\ref{truncatedMasterEquation}) subject to the boundary
conditions (\ref{boundaryCondition}) with $z=\infty$. For $\lambda \in {\mathbb R}$, define
$$
\widetilde{N}_L (\lambda) := \frac{\# \left \{ x \in (0,L):\, \psi(x,\lambda)=0 \right \}}{L}
$$
so that $L \widetilde{N}_L(\lambda)$ is the number of zeroes of the function
$$
x \mapsto \psi (x,\lambda)
$$
in $(0,L)$. Then $\lambda \mapsto L \widetilde{N}_L(\lambda)$ is a non-decreasing function and
$$
L \widetilde{N}_L (\lambda_n) = n\,.
$$
\label{sturmTheorem}
\end{theorem}
These facts are particular applications of {\em Sturm's oscillation and comparison theorems};
see Theorems 1.2 and 2.1
in Coddington \& Levinson (1955), Chapter 8. 

\begin{center}
\linethickness{2mm}
\color{light-grey}{\line(1,0){360}}
\end{center}
\begin{exercise}
Deduce from the theorem that, for every $\lambda \in {\mathbb R}$,
$$
\left | \widetilde{N}_L (\lambda) - N_L (\lambda) \right | \le \frac{1}{L}\,.
$$
\end{exercise}
\begin{center}
\linethickness{2mm}
\color{light-grey}{\line(1,0){360}}
\end{center}

The upshot is that we can study the large-$L$ limit of $N_L$ by counting the zeroes of
the function $x \mapsto \psi(x,\lambda)$ for $\lambda=k^2$ positive.
The evolution of this function between two neighbouring impurities has a very simple ``phase plane'' interpretation:
\begin{itemize}
\item On the interval $(x_j,x_{j+1})$ the evolution is free and the vector $(\psi, k\psi')$ undergoes a rotation of  random angle $k \ell_j$.
\item Across an impurity, the wave function is continuous but its derivative makes a  jump 
of random height $\psi'(x_j+0,\lambda)-\psi'(x_j-0,\lambda)=v_j\psi(x_j,\lambda)$.
\end{itemize}

\begin{figure}[htbp]
\vspace{7.5cm} 
\includegraphics{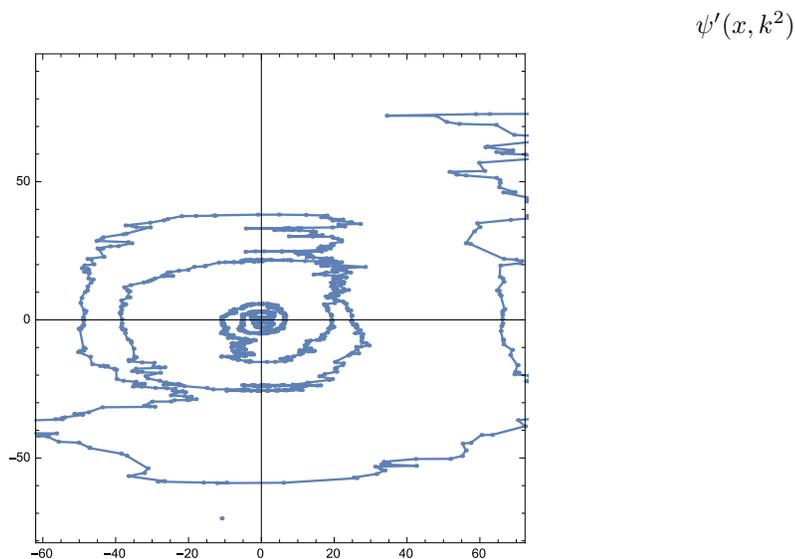}
\begin{picture}(0,0) 
\put(280,90){$\psi(x,k^2)$} 
\put(160,200){$\psi'(x,k^2)$} 
\end{picture}  
\caption{Typical phase-plane trajectory for the Schr\"odinger equation with a white noise potential and $E=k^2>0$.} 
\label{spiralFigure} 
\end{figure}

A typical trajectory is shown in  Figure \ref{spiralFigure}.  The figure illustrates two important 
features of the case $\lambda>0$: firstly, the trajectory winds around the origin and, secondly, it expands exponentially.
This motivates the parametrisation 
\begin{equation}
\frac{1}{\sqrt{\lambda}} \psi'(x,\lambda) + {\rm i}\,\psi(x,\lambda) =\varrho(x,\lambda) \,{\rm e}^{{\rm i} \theta(x,\lambda)} 
\label{parametrisation}
\end{equation}
of the solution $\psi(\cdot,\lambda)$ and its derivative. 
Mathematicians sometimes refer to the real variables $\varrho$ and $\theta$ as the {\em Pr\"{u}fer variables}; in terms of these variables, the differential equation
for $\psi$, namely
$$
-\psi'' + V(x) \psi = k^2 \psi\,,
$$
becomes
\begin{align}
&\theta'= k-\frac{V(x)}{k}\sin^2 \theta
\label{stoch1} \\
&\frac{\varrho'}{\varrho}= \frac{V(x)}{2k}\sin (2\theta)\,.
\label{stoch2}
\end{align}
Benderskii \& Pastur (1970) showed that, for a very large class
of processes $Q$,
\begin{equation}
\frac{\theta(L,\lambda)}{L}  \xrightarrow[L \rightarrow \infty]{} \pi N (\lambda)
\label{densityOfStates}
\end{equation}
almost surely.

The physical meaning of Equation (\ref{stoch1}) is quite clear: it describes the motion of a classical rotator under the action of a constant drift term (depending on the energy) and a random term which reduces or increases the drift depending on the sign of the potential. This is consistent with the heuristic idea that valleys in the random potential will create new states and thus increase the density of states. Equation (\ref{stoch2}) expresses the local growth rate of the wave function.
The same equations also arise in the context of dynamical systems perturbed by a noise--- a context in which their physical interpretation is even clearer. The general problem is to ascertain the effect of a stochastic perturbation 
on a time-independent integrable Hamiltonian  (Mallick and Marcq 2002; Hansel and Luciani 1989; Tessieri and Izrailev 2000). 
One the simplest questions concerns the stability of the system: what will happen for large times? 
Let us address this question in the case of 
a random-frequency oscillator with Hamiltonian
\begin{equation*}
H(q,p,t)=\frac{1}{2}{p}^2+\frac{1}{2} k^2 {q}^2-\frac{1}{2}{q}^2 V(t)\,.
\end{equation*}
The classical equation of motion coincides with the Schr\"{o}dinger equation after the substitution 
$(x,\psi) \rightarrow (t,q)$.  By disregarding the terms that do not
contribute to the large-time limit, it is easy to see that the Lyapunov exponent provides a measure of the asymptotic growth rate of  the system's energy:
\begin{equation*}
\gamma(k^2)=\lim_{t\rightarrow\infty}\frac{1}{t}\ln \left [ p^2 (t) + k^2 q^2(t) \right ]\,.
\end{equation*}
Thus, positivity of the Lyapunov exponent implies that the total energy stored in the system grows exponentially.
One can pursue the analysis a little further and view Equations (\ref{stoch1}-\ref{stoch2})  as equations satisfied by the action-angle variables $I=\sqrt\varrho$ and $\theta$ of the system, where
\begin{equation*}
\frac{{p}}{\sqrt{k}}:= \sqrt{2I} \cos\theta \;\;\text{and}\;\;
\sqrt{k} \,{q}:= \sqrt{2I}\sin\theta\,.
\end{equation*}
In the analysis of this problem, one expects that what really matters is not the total phase but the reduced phase
$$
\pi \left \{ \theta(x,\lambda) \right \} \in [0,\pi)
$$
where $\{ a \}$ denotes the fractional part of the number $a$.

\subsection{Riccati analysis: a qualitative picture}
\label{frischLloydAnalysisSubsection}
\begin{figure}[htbp]
\vspace{6cm} 
\includegraphics{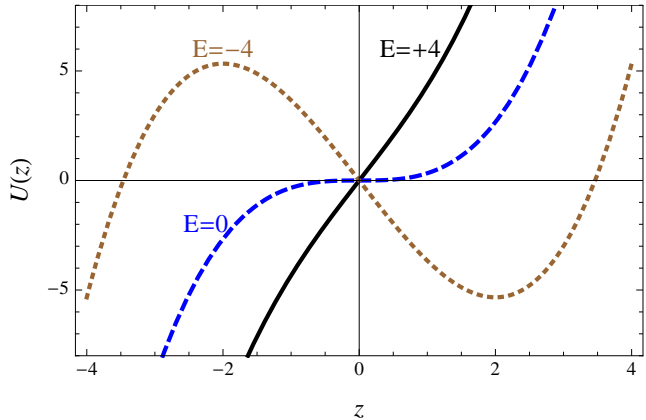}
\caption{The ``potential'' $U(z)$ associated with the unperturbed Riccati equation
$z' = -U'(z) = - (z^2+E)$.} 
\label{threePotentialsFigure} 
\end{figure}

It will be helpful in what follows to think of the spatial variable $x$ as ``time'', so that
the zeroes of $x \mapsto \psi(x,\lambda)$ then correspond to ``times'' at which the Riccati variable
$$
Z(x,\lambda) := \frac{\psi'(x,\lambda)}{\psi(x,\lambda)}
$$
blows up. For the Frisch--Lloyd model,  $Z(\cdot,\lambda)$ is a particular solution of
$$
Z' =  -Z^2 - \lambda + V\,.
$$
Between impurities, $V = 0$, and the Riccati equation 
is an autonomous system describing the motion of
a fictitious particle constrained to roll along the potential curve
$$
U(z) = \frac{z^3}{3} + \lambda z
$$
in such a way that its ``velocity'' at ``time'' $x$ and ``position'' $Z$ is given by the slope $-U'(z)$ (see
Figure \ref{threePotentialsFigure}):
\begin{equation}
Z' =-U'(Z ),\;\; x \notin \{ x_j \}_{j \in {\mathbb N}}\,.
\label{dynamicalSystem}
\end{equation}

When the particle hits an impurity, say $x_j$, the Ricatti variable makes a jump
$$
Z(x_j+)-Z(x_j-)=v_j\,.
$$
We may regard the occurence of these jumps
as a perturbation of the autonomous system (\ref{dynamicalSystem}), and the mean density of impurities as the perturbation parameter. 

Let us consider first the unperturbed system.
For $\lambda = k^2>0$, the system has no equilibrium point: the particle rolls down to $-\infty$, and re-appears
immediately at $+\infty$, reflecting the fact that the solution $\psi(\cdot,\lambda)$ of the corresponding
Schr\"{o}dinger equation has a zero at the ``time'' $x$ when the particle escapes to infinity. 
This behaviour of the Riccati variable indicates that every positive value of $\lambda$ belongs to the spectrum.

Let us now turn to the case $\lambda=-k^2 < 0$, $k>0$. In this case, the unperturbed system has an unstable equilibrium point at $-k$, and a stable equilibrium point at $k$.
Unless the particle starts at the unstable equilibrium, it must tend asymptotically to the stable equilibrium point. The fact that the particle cannot reach infinity more than once indicates that 
the spectrum lies entirely in ${\mathbb R}_+$. The solution of the Frisch--Lloyd equation is
$$
f = \delta (z-k)\,.
$$ 

Let us now consider how the occurence of jumps can affect the system. For $\lambda>0$, the jumps, as long as they are finite and infrequent (i.e. a small density of impurities), cannot prevent the particle from visiting $-\infty$ repeatedly; the system should therefore behave in much the same way as in the unperturbed case.
The situation for $\lambda = -k^2 <0$ is more complicated.
Roughly speaking,
{\em positive} jumps, i.e. discontinuous increases of $z$, enable
the particle to make excursions to the right of the stable equilibrium point $z=k$, but the particle can never overcome the
infinite barrier and so it rolls back down towards $k$. On the other hand,
{\em negative} jumps, i.e. discontinuous decreases of $z$, enable
the particle to make excursions to the left of $k$. If the jump is large enough, the particle can overcome the potential barrier at $-k$ and escape to
$-\infty$, raising the possibility that part of the spectrum of the Schr\"{o}dinger operator lies in ${\mathbb R}_-$. 

\subsection{The Rice formula}
\label{riceSubsection}
There remains to relate the zeroes
of $x \mapsto \psi(x,\lambda)$ to the complex Lyapunov exponent. 

In the absence of impurities, Equation ~(\ref{dynamicalSystem}) gives the velocity of the fictitious particle as a function of its position. Hence the time, say $\tau$, taken to go from $+\infty$ to $-\infty$  is
$$ 
\tau =-\int_{+\infty}^{-\infty}\frac{{\rm d} z}{z^2+k^2}=\frac{\pi}{k}\,.
$$
The fact that one ``particle" is transferred from $\infty$ to $-\infty$  in time $\tau$
means that there is a net current equal to
$k/\pi$. Of course,  $\tau$ is also the distance between two consecutive zeroes of $\psi(\cdot,\lambda)$. From the Sturm oscillation theorem, we deduce that the (free) integrated density of states per unit length is
$$
N(\lambda)=\frac{1}{\tau}=\frac{k}{\pi}\,.
$$ 

Now, the solution of the Frisch--Lloyd equation for $\lambda = k^2>0$ without impurities is the Cauchy density
$$
f(z) = \frac{k/\pi}{z^2+k^2}  \,.
$$
Hence
\begin{equation}
N (\lambda)=\lim_{|z|\to\infty}z^2 f(z)\,.
\label{rice}
\end{equation}
This relationship between the integrated density of states and the tail of the stationary density of the Riccati variable is a particular instance of the {\em Rice formula} (Rice 1944) and remains
valid in the presence of impurities. 

The Rice formula makes it clear that, in the spectrum, the stationary distribution of the Riccati variable does not have a mean. Nevertheless, the Cauchy principal value
$$
\dashint_{\mathbb R} z f(z)\,{\rm d} z := \lim_{a \rightarrow \infty} \int_{-a}^a z f(z) \,{\rm d} z
$$
exists, and it can be shown that
\begin{equation}
\text{Re} \,\Omega \left ( \lambda + {\rm i} 0 \right ) = \dashint_{\mathbb R} z f(z)\, {\rm d} z\,, \;\; \text{for $\lambda > 0$}  \,.
\label{cauchyFormula}
\end{equation}

\begin{center}
\linethickness{2mm}
\color{light-grey}{\line(1,0){360}}
\end{center}
\begin{exercise}
Verify the validity of Equation (\ref{cauchyFormula}) in the absence of impurities.
\label{cauchyExercise}
\end{exercise}
\begin{center}
\linethickness{2mm}
\color{light-grey}{\line(1,0){360}}
\end{center}

Let us revisit the concrete example discussed earlier in \S \ref{nieuwenhuizenSubsection},
and show how to exploit the Rice formula to connect $N$ with $\Omega$ when impurities are present.

For $\lambda = k^2 >0$, Equation (\ref{frischLloydDensityEquation}) must be replaced by
\begin{equation*}
(z^2+k^2) f(z) + \frac{1}{\ell} \,\int_{\text{supp} \rho} {\rm d} v_j \,\rho(v_j)\, \int_{z}^{z-v_j} {\rm d} t\,f(t) =  \frac{J}{\ell}
\end{equation*}
where $J$ is again the integration constant in Equation (\ref{integratedDysonSchmidtEquation}).
In particular, if we let $z \rightarrow \infty$, the Rice formula implies
\begin{equation}
J = \ell N\,.
\label{currentFormula}
\end{equation}
Hence $J$ measures the current of ``Riccati particles''.

The heuristic reasoning that follows borrows heavily from Halperin (1965).
Since, for $\lambda = k^2 >0$, the density $f$ is supported on the entire real line, we must now work
with the {\em Fourier transform}
\begin{equation}
F(q) := \int_{\mathbb R} f(z) \,{\rm e}^{-{\rm i} q z}\,{\rm d} z
\label{fourierTransform}
\end{equation}
of the unknown stationary density. For $\lambda \in {\mathbb C}_+$, $f$ decays sufficiently fast as $|z| \rightarrow \infty$ to ensure that
we can permute integration with respect to $z$ and differentiation with respect to $q$; this gives
\begin{equation}
{\rm i} F'(0) = \int_{\mathbb R} z f(z)\,{\rm d} z\;\;\text{for}\;\lambda \in {\mathbb C}_+\,.
\label{fourierTransformAtZero}
\end{equation}
By contrast, for $\lambda = k^2$, $f(z)$ decays like $1/z^2$ and so $F$ is not differentiable at $q=0$. Nevertheless, the following calculation makes clear that $F'(0+)$ and $F'(0-)$ do exist. Indeed, we have for the unknown transform $F$ a counterpart of Equation (\ref{laplaceEquation}):
\begin{equation}
\frac{{\rm d}^2 F}{{\rm d} q^2} + \frac{1}{\ell} \,{\mathbb E} \left ( \frac{ 1-{\rm e}^{-{\rm i} q v_j}}{{\rm i} q} \right ) \,F 
= \lambda F - 2\pi N(\lambda) \,\delta (q)\,.
\label{fourierEquation}
\end{equation}
Integrate this equation over $q$ from $-\varepsilon$ to $\varepsilon$ and let $\varepsilon \rightarrow 0+$; there comes
\begin{equation}
F'(0+)-F'(0-) = -2 \pi N (\lambda)\,.
\label{fourierJumpEquation}
\end{equation}
Now, from
$$
F(-q) = \overline{F(q)}\,, \;\; q \in {\mathbb R}\,,
$$
we deduce that
$$
{\rm i} F'(0-)  = \overline{{\rm i} F'(0+)}\,.
$$
Reporting this in Equation (\ref{fourierJumpEquation}), we find
\begin{equation}
\text{Im} \left [ {\rm i} F'(0\pm) \right ] = \mp \pi N (\lambda) \,.
\label{halperinImaginaryPart}
\end{equation}
On the other hand, 
\begin{equation}
\text{Re} \left [ {\rm i} F'(0+) \right ] = \text{Re} \left [ {\rm i} F'(0-) \right ] =  \dashint_{\mathbb R}  z f(z) \, {\rm d} z = \gamma(k^2)\,.
\label{halperinRealPart}
\end{equation}
Combining Equations (\ref{halperinImaginaryPart}) and (\ref{halperinRealPart}) yields
$$
{\rm i} F'(0 \pm) = \gamma(k^2) \mp {\rm i} \pi N (k^2)
$$
and when we compare this with Equation (\ref{fourierTransform}), we see that the equality
\begin{equation}
\Omega (k^2 \pm {\rm i} 0) = {\rm i} F'(0 \pm)
\label{halperinFunction}
\end{equation}
is plausible. In particular,
$$
-\frac{1}{\pi} \,\text{Im}\,\Omega( k^2+{\rm i} 0) = N (k^2)\,.
$$

\section{The white noise limit}
\label{whiteNoiseSection}

We have seen that the Frisch-Lloyd model can sometimes be ``solved'' explicitly when the integral equation for the stationary density of the Riccati variable can be transformed into an ordinary differential equation. The question of the extent to which this can be done for an arbitrary discrete model has been the subject of recent investigations that we will not discuss here. Our goal is more modest: we shall focus on a particular limit known as the {\em white noise limit}. It describes situations in which the impurities are densely packed along the real
line whilst the variance of the coupling constants is small--- a scenario which suggests a description in terms of a continuous model. We will show that the limit model is one in which the potential is a Gaussian process with delta correlations. 

\subsection{L\'{e}vy processes}
\label{levySubsection}

Let us write, as before,
\begin{equation}
V(x) = Q' (x) := \sum_{j=1}^\infty v_j \delta (x-x_j) \,.
\label{impurityPotential}
\end{equation}
and consider the particular case where the spacings $\ell_j$ between successive impurities
are exponentially distributed, independent random variables, and where the $v_j$ are also identically distributed
random variables with a common probability density $\rho$. This defines a random process 
\begin{equation}
Q(x) := \int_0^x V(t) \,{\rm d} t\,.
\label{processQ}
\end{equation}
The simplest case corresponds to taking all the $v_j$ equal to one.
Then
$$
Q(x) = n(x) := \# \left \{ j \in {\mathbb N} : \, x_j < x \right \}
$$
is the number of impurities contained in the interval $(0,x)$. For our choice of $\ell_j$, this 
is the familiar {\em Poisson process} of intensity $1/\ell$, which has
the following properties (Feller 1971):
\begin{enumerate}
\item The process starts at $0$, i.e.
$$
Q(0)=0\,.
$$
\item For every realisation of the process, the function 
$$
x \mapsto Q(x)
$$ 
is right-continuous, and the limit
$$
\lim_{y \rightarrow x-} Q(y)
$$
exists for every $x > 0$.
\item The increments are stationary, i.e. for every $x,\,y \ge 0$,
$$
Q(x+y) - Q(x) \overset{\text{law}}{=} Q(y)\,.
$$
\item The increments are independent, i.e. for every $0 \le u < x \le  y < z$,
$$
Q(x) - Q(u) \;\;\text{and}\;\;Q(z)-Q(y)
$$
are independent.
\end{enumerate}
A process with these properties is called a {\em L\'{e}vy process}.  For other distributions of the $v_j$, 
the impurity potential (\ref{impurityPotential}) can be expressed in the form
\begin{equation}
Q(x) = \sum_{j=1}^{n(x)} v_j\,.
\label{compoundPoissonProcess}
\end{equation}
$Q$ is then called a {\em compound Poisson Process} and it retains Properties (1-4).
There are other L\'{e}vy processes which are {\em not} of the form (\ref{compoundPoissonProcess});
see (Applebaum 2004). The best-known example is {\em Brownian motion}:
this is a continuous Gaussian stochastic process on the positive half-line 
such that, for every $x,\,y \ge 0$, there holds
$$
{\mathbb E} \left [ Q(x) \right ] = 0 \;\;\text{and}\;\; {\mathbb E} \left [ Q(x) Q(y) \right ] = \min \{ x,\,y\}\,.
$$

We have thus embedded our impurity potential in a larger class which includes in particular
the Gaussian white noise potential
\begin{equation}
V := \sqrt{\sigma} B'
\label{whiteNoisePotential}
\end{equation}
with $\sigma > 0$.  There remains to explain how, and in what sense, $V$ may be approximated by potentials of the form (\ref{compoundPoissonProcess}).

By exploiting the properties satisfied by every L\'{e}vy process $Q$, it may be shown that
$$
{\mathbb E} \left  [ {\rm e}^{{\rm i} \theta Q(x)} \right ] = \exp \left [ x \Lambda (\theta) \right ]
$$
where
\begin{equation}
\Lambda (\theta) := \lim_{h \rightarrow 0} {\mathbb E} \left [ \frac{{\rm e}^{{\rm i} \theta Q(h)} -1}{h} \right ]\,.
\label{levyExponentDefinition}
\end{equation}
$\Lambda$ is called the {\em L\'{e}vy exponent} of $Q$ and completely characterises it.

\begin{example}
Let us compute the L\'{e}vy exponent of the compound Poisson process corresponding to the impurity potential
(\ref{impurityPotential}) or, what is the same, (\ref{compoundPoissonProcess}). It is easy to show that
$$
{\mathbb P} \left ( n(x) = j \right ) = {\rm e}^{-x/\ell} \frac{(x/\ell)^j}{j!}\,.
$$
Then
\begin{multline}
\notag
{\mathbb E} \left ( {\rm e}^{{\rm i} \theta Q(x)} \right ) = \sum_{j=0}^\infty {\mathbb E} \left ( {\rm e}^{{\rm i} \theta Q(x)} \bigl | \,n(x) = j \right )
{\mathbb P} \left ( n(x) = j \right ) \\
= \sum_{j=0}^\infty {\mathbb E} \left ( {\rm e}^{{\rm i} \theta \sum_{i=1}^j v_i} \right ) 
{\rm e}^{-x/\ell} \frac{(x/\ell)^j}{j!}  
= {\rm e}^{-x/\ell} \sum_{j=0}^\infty {\mathbb E} \left ( {\rm e}^{{\rm i} \theta v_1} \right )^j 
 \frac{(x/\ell)^j}{j!} \\
= \exp \left \{ x \frac{{\mathbb E} \left ( {\rm e}^{{\rm i} \theta v_1} \right ) -1}{\ell} \right \}\,.
\end{multline}
Hence
$$
\Lambda (\theta) = \frac{{\mathbb E} \left ( {\rm e}^{{\rm i} \theta v_1 } \right )-1}{\ell}\,.
$$
\label{levyExponentExample1}
\end{example}

\begin{center}
\linethickness{2mm}
\color{light-grey}{\line(1,0){360}}
\end{center}
\begin{exercise}
Consider the special case where the probability density of the $v_j$ is given by
\begin{equation}
\rho(x)= \frac{a}{2} {\rm e}^{- a |x|}\,,\;\; a > 0\,.
\label{symmetricDensity}
\end{equation}
\begin{enumerate}[label=(\alph*)]
\item What are the mean and the variance of the $v_j$?
\item Work out the L\'{e}vy exponent.
\item Let $\sigma > 0$. Show that
\begin{equation}
\lim_{\stackrel{\ell \rightarrow 0,\, a \rightarrow \infty}{\ell a^2 = 2/\sigma}} \Lambda (\theta) = -\frac{\sigma \theta^2}{2}
\label{symmetricLimit}
\end{equation}
\item Show that the right-hand side of (\ref{symmetricLimit}) is the L\'{e}vy exponent of $\sqrt{\sigma} B$, where
$B$ is a standard Brownian motion.

{\em Help:} 
Use the fact that, for every $x > 0$, $B(x)$ is a Gaussian random variable of mean $0$ and variance $x$.
\end{enumerate}
\label{continuumExercise}
\end{exercise}
\begin{center}
\linethickness{2mm}
\color{light-grey}{\line(1,0){360}}
\end{center}

For the particular impurity model in this exercise, we thus have
\begin{equation}
V {\rightharpoonup} \,\sqrt{\sigma} B'\,.
\label{weakLimit}
\end{equation}
This weak limit is in the sense of convergence of the L\'{e}vy exponent. Although in the exercise we considered a specific choice of distribution for the coupling constants, the conclusion
holds for a large class of models; what really matters is the scaling condition that ${\mathbb E}(v_j^2)/ \ell$ tend to a non-zero finite value.

\subsection{The Lyapunov exponent}
\label{whiteNoiseLyapunovSubsection}

Let us now turn to the calculation of the complex Lyapunov exponent of the continuum limit, We begin by remarking that
Equation (\ref{fourierEquation}) for the Fourier transform of the invariant probability density $f(z)$ of the Riccati variable may be expressed in terms of the L\'{e}vy exponent as follows:
\begin{equation*}
F'' - \frac{\Lambda (- q )}{{\rm i} q}  \,F 
= k^2 F - 2\pi N(k^2) \,\delta (q)\,.
\end{equation*}
In particular, for the white noise potential $V= \sqrt{\sigma} B'$, this becomes
\begin{equation}
F'' - {\rm i} \frac{\sigma}{2} q F = k^2 F\,,\; q > 0\,.
\label{halperinEquation}
\end{equation}
This equation can easily be reduced to the Airy equation (Abramowitz and Stegun 1964). The Fourier transform of $f$ is proportional to the solution that decays to zero as $q \rightarrow \infty$. Hence
$$
F(q) = c \left [ \text{Ai} \left ( -k^2/\xi^2-{\rm i} \xi q \right ) - {\rm i} \,\text{Bi} \left ( -k^2/\xi^2-{\rm i} \xi q \right ) \right ]\,, \;\; \xi :=  \left ( \sigma/2 \right )^\frac{1}{3}\,.
$$
By the same reasoning as in \S \ref{riceSubsection}, we eventually obtain
\begin{equation}
\Omega (k^2+{\rm i}0) = \xi\, \frac{\text{Ai}' \left ( - k^2/\xi^2 \right ) - {\rm i} \,\text{Bi}' \left ( - k^2/\xi^2 \right )}{\text{Ai} \left ( - k^2/\xi^2 \right ) - {\rm i} \,\text{Bi} \left ( - k^2/\xi^2 \right )}\,.
\label{continuumLyapunov}
\end{equation}
This is a minor extension of a calculation first performed by Halperin (1965).

\begin{center}
\linethickness{2mm}
\color{light-grey}{\line(1,0){360}}
\end{center}
\begin{exercise}
Deduce that, for the white noise potential $V = \sqrt{\sigma} B'$,
the integrated density of states is given by
$$
N (k^2) = \frac{\xi/\pi^2}{\text{Ai}^2 \left ( - k^2/\xi^2 \right ) +\text{Bi}^2 \left ( - k^2/\xi^2 \right )}
\,, \;\; \xi :=  \left ( \sigma/2 \right )^\frac{1}{3}\,.
$$ 
\label{halperinExercise}
\end{exercise}
\begin{center}
\linethickness{2mm}
\color{light-grey}{\line(1,0){360}}
\end{center}

One can also find a more elementary expression for $N$ by working with the stationary density
of the Riccati variable. Indeed, applying the inverse Fourier transform to Equation (\ref{halperinEquation}) yields
$$
\frac{\sigma}{2} f'(z) + (z^2+k^2) f(z) = N(k^2)\,.
$$
Hence
\begin{equation}
f(z) = \frac{2 N(k^2)}{\sigma} \exp \left [ -\frac{2}{\sigma} \left ( \frac{z^3}{3} + k^2 z \right ) \right ]
\int_{-\infty}^z {\rm d} t \exp \left [ \frac{2}{\sigma} \left ( \frac{t^3}{3} + k^2 t \right ) \right ]\,.
\label{halperinDensity}
\end{equation}

\begin{center}
\linethickness{2mm}
\color{light-grey}{\line(1,0){360}}
\end{center}
\begin{exercise}
Deduce the integral formula
\begin{equation}
\frac{1}{N (k^2)} = \sqrt{2 \pi/\sigma} \int_0^\infty \frac{{\rm d} u}{\sqrt{u}} \exp \left [ -\frac{2}{\sigma}  \left ( \frac{u^3}{12}+k^2 u \right )
\right ]  \,.
\label{halperinFormula}
\end{equation}
{\em Help:} Integrate Equation (\ref{halperinDensity}) over $z$. This results in two nested integrals; by making a judicious substitution for one of the variables, one integral can be done explicitly.  
\label{halperinDensityExercise}
\end{exercise}
\begin{center}
\linethickness{2mm}
\color{light-grey}{\line(1,0){360}}
\end{center}

\subsection{Riccati analysis}
\label{whiteNoiseRiccatiSubsection}
Let us indicate briefly how this result can also be derived by a Riccati analysis in the spirit of \S \ref{frischLloydAnalysisSubsection}. Sturm's oscillation theorem states that the density of states is obtained by counting the number of zeroes of the wave function, which is also the number of times the process $\{Z(x)\}_{x \ge 0}$ goes to $-\infty$. Denote  these
random times by $\tau_j$, so that
\begin{align*}
&\tau_1 =\inf \{ x>0 : Z(x)=-\infty\} \\
&\tau_2=\inf \{ x> \tau_1 : Z(x)=-\infty\} \\
&\ldots \\
&\tau_n=\inf \{ x> \tau_{n-1} : Z(x)=-\infty\}\,.
\end{align*}
The integrated density of states is
$$
N(E)=\lim_{L\to\infty}\frac{1}{L}\#\{j :\,\tau_j \le L\}=\lim_{n\to\infty}\frac{n}{\tau_n}\,.
$$
Therefore, if we assume that the process $\{Z(x)\}_{x \ge 0}$ is ergodic, one finds
\begin{equation}
N(E)=\frac{1}{{\mathbb E}_\infty( \tau_1 )}
\label{mcKeanFormula}
\end{equation}
where
$$
{\mathbb E}_z  (\tau_1) = {\mathbb E} \left ( \tau_1 \Bigl | \,Z(0) = z \right )\,.
$$
It may be shown by using the tools of stochastic calculus that the Laplace transform 
\begin{equation*}
h_p(z )  := {\mathbb E}_z \left ( {\rm e}^{-p \tau_1} \right )
\end{equation*}
is the recessive solution of the equation
$$
\left [ \frac{\sigma}{2} \frac{{\rm d}^2}{{\rm d} z^2} - (z^2+k^2) \frac{{\rm d}}{{\rm d} z} \right ] u = p\, u
$$
that satisfies the condition $u(z)=1$ as $z \rightarrow -\infty$. By using the obvious integrating factor,
we can reformulate the differential equation
as an integral equation and incorporate the condition at $-\infty$:
\begin{equation}
h_p(z)=1-\frac{2p}{\sigma}\int_{-\infty}^{z}{\rm d} x \, {\rm e}^{\frac{2}{\sigma} \left ( \frac{x^3}{3}+k^2x \right )} \int_{x}^{\infty}{\rm d} y\, {\rm e}^{-\frac{2}{\sigma} \left (\frac{y^3}{3} + k^2y\right )}\,h_p(y)\,.
\label{Halp2}
\end{equation}

\begin{center}
\linethickness{2mm}
\color{light-grey}{\line(1,0){360}}
\end{center}
\begin{exercise}
Use Equation (\ref{Halp2}) to recover Formula (\ref{halperinFormula}).
\end{exercise}
\begin{center}
\linethickness{2mm}
\color{light-grey}{\line(1,0){360}}
\end{center}

\section{Lifshitz tails and Lifshitz singularities}
\label{lifshitsSection}

The density of states and the Lyapunov exponent often display interesting behaviour at the bottom of the spectrum. For instance if the potential is a white noise and is therefore unbounded one can probe the region  $E\rightarrow -\infty$ by a saddle approximation of (\ref{halperinFormula}):
 \begin{equation}
 \ln N (E)\underset{E\to -\infty}{\sim} -\frac{8}{3\sigma}|E|^{3/2}
 \label{saddleApproximation}
 \end{equation}
This limit behaviour is called a {\em Lifshitz tail}. Because the behaviour is not analytic in $E$, one says that
$N$ has a {\em Lifshitz singularity} at $-\infty$.
It may be explained by a very simple physical argument that involves the Riccati process (Jona--Lasinio 1983). As shown in \S \ref{frischLloydAnalysisSubsection}, for $E=-k^2 <0$, the effective potential $U(z)$ develops a local minimum at $ z=-k$ and a local maximum at $z=k$; this creates a potential barrier of height
$$\Delta U= U(-k)-U(k)= \frac{4k^3}{3}\,.$$
 The expected hitting time of $-\infty$ is essentially given by the time that the particle takes 
 to overcome the potential barrier. The mean first passage time, estimated by the Arrhenius formula, is
$$
{\mathbb E}_{\infty}(\tau_1)= {\rm e}^{\frac{2\Delta U}{\sigma}}
$$
and so (\ref{saddleApproximation}) follows. Lifshitz singularities occur in several disordered systems and have been extensively studied in the literature. Let us briefly review some interesting cases.

\subsection{Short-range, repulsive potentials} Then 
$$ 
\ln N(E) \sim  -\frac{c}{\sqrt{E}} \;\;\text{as $E \rightarrow 0+$}\,.
$$
Again, this non-analytic behaviour has a simple explanation, 
due to Lifshitz: Low-energy modes may appear when there are large regions of space that are free of impurities, in which the particle can move freely. Such configurations have an exponentially small probability. For instance, if the impurities are uniformly distributed with mean spacing $\ell$, then the probability of finding an impurity-free region of size $L$ is ${\rm e}^{- L/\ell}$. Since the lowest-energy mode is of order $k=\pi/L$ this gives 
$$
\ln N(E) \sim -\frac{\pi}{\ell \sqrt E}\;\;\text{as $E \rightarrow 0+$}\,.
$$
This argument generalises to $d$ dimensions; the exponent of $E$ becomes $-d/2$--- a result that may also be derived via instanton techniques. 
\smallskip

\subsection{Strings of Type I} 
For Dyson's strings, the behaviour at the bottom of the
spectrum depends on the type: The Type II string of
Exercise \ref{secondLetacExercise}
exhibits a square-root behaviour; by contrast,
Dyson showed that the string of Type I discussed in \S \ref{typeISubsection} is such that
$$
\ln N (\lambda) \sim -2 \ln \left | \ln \lambda \right | \;\;\text{as $\lambda \rightarrow 0+$}\,.
$$ 
Such singularities are called {\em Dyson singularities}.

\begin{center}
\linethickness{2mm}
\color{light-grey}{\line(1,0){360}}
\end{center}
\begin{exercise}
Consider a random string of Type I, with a Dirichlet condition at $x=0$ (i.e. $\alpha=\pi/2$), such that
$$
c_j \sim \text{Gamma} (p,q)\,.
$$
Show that, for $\lambda < 0$,
$$
{\mathbb E} \left [ w(\lambda) \right ] = -\lambda \frac{\Psi '(p,1;-\lambda/q)}{\Psi (p,1;-\lambda/q)}\,.
$$
\label{typeIExercise2}
\end{exercise}
\begin{center}
\linethickness{2mm}
\color{light-grey}{\line(1,0){360}}
\end{center}

Let us use this to compute ${\mathbb E} (\sigma')$. Analytic continuation gives, for $\lambda > 0$,
$$
\frac{{\mathbb E} \left [ w(\lambda+{\rm i} 0) \right ]}{\lambda}
= -\frac{\Psi '(p,1;-\lambda/q-{\rm i}0)}{\Psi (p,1;-\lambda/q-{\rm i}0)}
= -\frac{\Psi '(p,1;-\lambda/q-{\rm i} 0) \Psi (p,1;-\lambda/q+{\rm i}0)}{\left | \Psi (p,1;-\lambda/q-{\rm i}0) \right |^2}\,.
$$
Therefore
\begin{multline}
\notag
-\frac{2 {\rm i} \,\text{Im} \,w(\lambda + {\rm i} 0)}{\lambda} \\
= \frac{\Psi '(p,1;-\lambda/q-{\rm i} 0) \Psi (p,1;-\lambda/q+{\rm i}0)-\Psi '(p,1;-\lambda/q+{\rm i} 0) \Psi (p,1;-\lambda/q-{\rm i}0)}{\left | \Psi (p,1;-\lambda/q-{\rm i}0) \right |^2}\,.
\end{multline}
Ismail \& Kelker (1979) worked out a simple expression for the numerator of the right-hand side; their Equation (3.5) says that this numerator equals
$$
- \frac{2 \pi {\rm i}}{\lambda} \,\frac{q \,{\rm e}^{-\lambda/q}}{\Gamma(p)^2}\,.
$$
Hence
$$
{\mathbb E} \left [ \sigma'(\lambda) \right ] = \frac{q {\rm e}^{-\lambda/q}}{\Gamma(p)^2 \left | \Psi (p,1;-\lambda/q-{\rm i}0) \right |^2}\,.
$$
It follows in particular that
$$
{\mathbb E} \left [ \sigma'(\lambda) \right ]  \sim \frac{q}{\ln^2 (\lambda/q)}\;\;\text{as $\lambda \rightarrow 0+$}\,.
$$

\subsection{Supersymmetric potentials} A potential of the form
\begin{equation*}
V(x)= W'(x) + W^2(x)
\label{superSymmetry}
\end{equation*}
is called {\em supersymmetric} and
$W(x)$ is called the {\em superpotential} .  For such potentials, the spectral measure of the Schr\"{o}dinger
Hamiltonian is always supported on $[0,\infty)$. When $W(x)$ is random, the Lifshitz tails
can be of the form
 $$
 \ln N(E) \sim c \ln E \;\;\text{as $E \rightarrow 0+$}
 $$ 
or feature a Dyson singularity. A physical picture that accounts for such behaviours is presented in (Comtet {\em et al.} 1995). It turns out that the mechanism is just the opposite of the Lifshitz mechanism described earlier.  
Instead of being localised  in  impurity-free regions, the low-energy states are localised near the impurities.
We note incidentally  that the Schr\"odinger equation with a supersymmetric potential can be recast 
as the first-order system
\begin{equation*}
- \psi' + W \psi = \sqrt{E} \,\phi \quad \text{and} \quad \phi' + W \phi = \sqrt{E} \,\psi\,.
\end{equation*}
These equations may be viewed as a degenerate form of the Dirac equation with the Hamiltonian 
\begin{equation*}
{\mathscr H} := 
\begin{pmatrix}
{\mathscr O} & {\mathscr Q} \\
{\mathscr Q}^{\dag} & {\mathscr O}
\end{pmatrix} 
\end{equation*}
Where ${\mathscr O}$ denotes the zero operator and ${\mathscr Q}$ is a first-order differential operator. Hamiltonians with such a structure are called {\em chiral}. They obey the symmetry condition
$$
{\mathscr H}=-\sigma_3 {\mathscr H} \sigma_3\,. $$
It implies that positive and negative energy levels appear in pairs $\pm E$; if $\psi_E$ 
is an eigenstate of energy $E$ then $\sigma_3 \psi_E $ is an eigenstate of energy $-E$. 
The presence of this symmetry  places severe constraints on the localisation properties of the system. 
The one-dimensional supersymmetric Hamiltonian is thus an interesting toy model that
illustrates the importance of symmetry considerations in disordered systems (Altland and Zirnbauer 1997).

\subsection{Further reading}
An introduction to supersymmetric quantum mechanics  and a comprehensive review of its applications in quantum and statistical physics can be found in the book by Junker (1996). These models are relevant in several context of condensed matter physics, including one dimensional disordered semiconductors (Ovchinnikov and \'{E}rikhman 1977), random spin chains and organic conductors;
see Comtet \& Texier (1998) for a review. They have also found interesting applications in the context of diffusion in a random environment (Bouchaud {\em et al.} 1990; Grabsch {\em et al.} 2014; Le Doussal 2009).  A table summarising the low-energy behaviours  of such models is given in Grabsch {\em et al.} (2014). Further information on their spectral properties can be found in Comtet {\em et al.} (2011)\,.

\section{Distribution of the ground state energy and statistics of energy levels}
\label{groundStateSection}
We have shown that the correct tail behaviour of the integrated density of states
may be obtained by using a simple analogy with a physical activation process.
The same analogy can also be used to investigate finer properties of the spectrum, such as the distribution of the ground state energy and the statistics of energy levels. In order to give a precise meaning to this statement, consider the truncated spectral problem 
$$ 
{\mathscr H} \psi =E \,\psi \,, \;\; 0 < x < L\,,
$$
with the Dirichlet boundary conditions 
$$
\psi (0)=\psi(L)=0\,.
$$
In particular we would like to compute the probability distribution of the ground state energy $E_0$.
As before, this problem can be reduced to studying the zeroes of the particular solution
$\psi (\cdot,E)$ of the truncated problem
that satisfies the initial conditions $\psi(0,E)=0, \psi'(0,E)=1$.
Denoting by $\tau_1$  the first strictly positive zero of $ \psi(\cdot,E)$, Sturm's oscillation theorem implies that the event $E_0>E$ is the same as
the event $\tau_1>L$. Hence
$$ \mathbb{P}(E_0>E)=\mathbb{P}(\ell>L)$$
and the problem of computing the statistics of $E_0$ amounts to solving a {\em first passage problem} for the Riccati process $\{Z(x)\}_{x \ge 0}$ started at infinity (Grenkova {\em et al.} 1983) .

Following Texier (2000), let us show how to compute the tail of this distribution by using the activation analogy.
As discussed earlier, the expected exit time from the potential barrier is related to the integrated density of states 
via
$$\tau := {\mathbb E}_\infty (\tau_1)=\frac{1}{N(E)}$$
We require--- not just the mean--- but rather the whole distribution of the time of escape from the well. In the 
weak-disorder limit $E\sigma^{-2/3}\rightarrow -\infty$, the probability density, say $p$, of $\tau_1$  is given by 
$$
p(t)=\frac{1}{\tau}{\rm e}^{-\frac{t}{\tau}}\,.
$$
Therefore the ground state energy distribution is given by
\begin{equation}
 \mathbb{P}(E_0>E)= \mathbb{P}(\tau_1>L)=\int_L^{\infty} p(t) \,{\rm d} t= {\rm e}^{-LN(E)}\,.
 \label{weakDisorderEstimate}
\end{equation}
When the size $L$ of the system is very large, the ground state energy $E_0$--- properly centered and rescaled--- will have a limiting law. 
Take for example the case of the white noise potential $V = \sqrt{\sigma} B'$ and set for simplicity $\sigma=1$ so that the tail of the density of states is given by
\begin{equation}
N(E)= \frac{\sqrt {-E}}{\pi} {\rm e}^{-\frac{8}{3} (-E)^{3/2}}
\end{equation}
One expects that Equation (\ref{weakDisorderEstimate}) will give the correct tail of the true probability distribution 
in the limit 
$L\rightarrow \infty$.
After some algebra, one gets
$$
\lim_{L\rightarrow\infty}\mathbb{P}\left[\frac{E_0+({\frac{3}{8}\log \frac{L}{\pi}})^{2/3}}{(24\log L)^{-1/3}} < x \right]= 1-{\rm e} ^{-{\rm e}^ x}\,.
$$
This result, derived for the case of a white noise potential in McKean (1994), 
is in fact discussed in greater generality 
in an earlier paper  (Grenkova {\em et al.} 1983).
As pointed out in Texier (2000), it expresses the fact that the distribution of $E_0$--- properly centered and rescaled---
 is a {\em Gumbel law}. This distribution belongs to one of the three classes of  extreme value statistics characterising the minimum or the maximum of a set of independent, identically distributed random variables.

For the Frisch-Lloyd model where the $x_n$ are the points of a Poisson process and the $v_n$ 
independent, identically distributed random variables with an exponential distribution (Grenkova {\em et al.} 1983),
and for its white noise limit (Texier 2000),
the analysis can be pushed a little further and generalised to excited states. 
One finds
$$\mathbb{P}[E_{n-1}<E<E_{n}]= \mathbb{P}[\tau_1+\tau_2+...+\tau_n <L< \tau_1+\tau_2+....+\tau_{n+1}]$$
where the $\tau_j$ are independent and have the same distribution as $\tau_1$. Thus, for $n=1,2,\ldots$\,,
\begin{displaymath}
\mathbb{P}[E_{n-1}<E<E_{n}]= \frac{\left [ LN(E) \right ]^n}{n!} {\rm e}^{-LN(E)} 
\end{displaymath}
and
\begin{displaymath}
\lim_{L\rightarrow\infty}\mathbb{P}\left[\frac{E_{n-1}-f_{n-1}(L)}{\sigma_{n-1} (L)}\in dx\right]= \frac{n^{n-1/2}}{(n-1)!} \exp\left( x\sqrt n-n {\rm e}^{x/\sqrt n}\right) {\rm d} x\end{displaymath}
for some functions $f_n$ and $\sigma_n$ that can be worked out explicitly.
This result expresses the fact that the energy levels are uncorrelated (Grenkova {\em et al.} 1983; Mol\u{c}anov 1981), and this is expected to hold whenever  the quantum states are localised. When delocalization occurs at the bottom of the spectrum, as it does for the supersymmetric disordered Hamiltonian, the first energy levels  are correlated and the energy levels do not obey Gumbel
laws (Texier 2000; Texier and Hagendorf 2010).

\section{Scattering and hyperbolic geometry}
\label{scatteringSection}

So far, our discussion of impurity models has focused on the spectral properties of the master equation
(\ref{masterEquation}) and of its truncated version (\ref{truncatedMasterEquation}). Another problem of physical interest concerns the effect of disorder on the transmission of a plane wave through a sample. This problem may be stated as follows: Given
$\lambda = k^2> 0$ and $L>0$, find complex numbers $R_L$ and $T_L$ such that there is a solution $\psi_L(x)$ of Equation (\ref{truncatedMasterEquation}) satisfying
\begin{equation}
\psi_L(x) = \begin{cases}
{\rm e}^{{\rm i} k x} + R_L \,{\rm e}^{-{\rm i} k x} & \text{if $x \le 0$} \\
T_L \,{\rm e}^{{\rm i} k x} & \text{if $x \ge L$}
\end{cases}\,.
\label{scatteringConditions}
\end{equation}
The numbers $R_L$ and $T_L$ are called the reflexion coefficient and the transmission coefficient respectively.

Our aim in this section will be to determine how the transmission coefficient changes as the length $L$ increases. 

\begin{remark}
This problem makes sense for both the Frisch--Lloyd and Dyson models. For the string equation, however, it
is more natural to consider the generalised version in Exercise \ref{generalisedStringExercise} with $\mu=1$.
\label{scatteringRemark}
\end{remark}

\begin{center}
\linethickness{2mm}
\color{light-grey}{\line(1,0){360}}
\end{center}
\begin{exercise}
Show that
$$
\left | R_L \right |^2 + \left | T_L \right |^2 = 1\,.
$$
{\em Help:} Use Green's identity.
\label{reflexionTransmissionExercise}
\end{exercise}
\begin{center}
\linethickness{2mm}
\color{light-grey}{\line(1,0){360}}
\end{center}

Recall that the general solution of the truncated equation (\ref{truncatedMasterEquation}) is 
$$
\begin{pmatrix}
\psi'(x_{n+1}-) \\
 \psi (x_{n+1})
\end{pmatrix}
= \Pi_n  A_0 \begin{pmatrix} 
\psi'(0-) \\
k \psi (0)
\end{pmatrix}
$$
where the matrices $A_j$ are those given in \S \ref{impuritySection}.
Let now 
$$
L = x_{n+1}\;\;\text{and}\;\;k=1\,.
$$
(This restriction on $k$ is merely for convenience and entails no real loss of generality.)
By working with the Riccati variable
\begin{equation}
Z_L = \frac{\psi_L'}{\psi_L}
\label{scatteringRiccatiVariable}
\end{equation}
we deduce easily that
\begin{equation}
R_{L}  = - \frac{{\mathcal X}_{n+1}({\rm i})-{\rm i}}{{\mathcal X}_{n+1}({\rm i})+{\rm i}}\,. 
\label{reflexionFormula}
\end{equation}
where
\begin{equation}
{\mathcal X}_{n+1} := {\mathcal A}_0^{-1} \circ {\mathcal A}_1^{-1} \circ \cdots \circ {\mathcal A}_n^{-1} \,.
\label{riccatiScattering}
\end{equation}
The sequence $\{ {\mathcal X} _n\}_{n \in {\mathbb N}}$ defined by this last equation
satisfies the recurrence relation
\begin{equation}
{\mathcal X}_{n+1} = {\mathcal X}_{n} \circ {\mathcal A}_{n}^{-1} \,.
\label{randomWalk}
\end{equation}
In particular, when the $A_n$ are independent and identically distributed, the sequence
is a random walk in the group of linear fractional transformations 
on the upper half-plane 
$$
{\mathbb H} := \left \{ x + {\rm i} y :\, x \in {\mathbb R},\, y > 0 \right \}
$$ 
starting from the identity element.

\begin{center}
\linethickness{2mm}
\color{light-grey}{\line(1,0){360}}
\end{center}
\begin{exercise}
Let $A \in \text{SL}(2,{\mathbb R})$. Show that 
$$
{\mathcal A} \left ( {\mathbb H} \right ) \subseteq {\mathbb H}\,.
$$
\label{hyperbolicExercise}
\end{exercise}
\begin{center}
\linethickness{2mm}
\color{light-grey}{\line(1,0){360}}
\end{center}

\subsection{Hyperbolic geometry}
\label{hyperbolicSubsection}

Let us step back from impurity models for a moment and consider the equation satisfied by 
the Riccati variable (\ref{scatteringRiccatiVariable}):
\begin{equation}
Z' =- Z^2+Q'- M',\, \;\; 0 \le x < L\,,
\label{hyperbolicRiccatiProcess}
\end{equation}
where, for convenience, we assume that $Q$ and $M$ are differentiable in the usual sense. Put
$$
Z = X + {\rm i} Y
$$
so that
$$
X' = Q' - M' +Y^2 - X^2 \;\;\text{and}\;\; Y' = -2  X Y\,.
$$
There are two possibilities: either $Y$ is identically zero or else it is always of the same sign. The case of interest for our purposes is obviously $Y>0$.
Then $\{ Z(x) : x \ge 0\}$ is a curve contained in the upper half-plane. This is consistent with the result of Exercise 
\ref{hyperbolicExercise}.

Further insight into the nature of the Riccati flow may be gained by renaming the independent variable $t$ and setting
$$
q := X \;\;\text{and}\;\; p := 1/Y\,.
$$
Then the Riccati equation is equivalent to the real Hamiltonian system
$$
\dot{p} = -\frac{\partial H}{\partial q} \;\;\text{and}\;\;\dot{q} = \frac{\partial H}{\partial p}
$$
where 
\begin{equation}
H (q,p) :=  \left ( Q' - M' -  q^2 \right ) p -\frac{1}{p} \,.
\label{riccatiHamiltonian}
\end{equation}
In particular, by Liouville's theorem, the Hamiltonian flow in phase space is {\em incompressible}. In terms of the original Riccati variable,
the volume element is
\begin{equation}
{\rm d} p \,{\rm d} q = \frac{{\rm d} X\, {\rm d} Y}{Y^2}
\label{hyperbolicVolumeElement}
\end{equation}
and this is conserved by the Riccati flow.

It turns out that this is also precisely the infinitesimal volume associated with Poincar\'{e}'s half-plane model of two-dimensional
hyperbolic geometry. In this model, the length, say $l_Z$, of a path
$$
Z = X+ {\rm i} Y : [0,1] \rightarrow {\mathbb H}
$$ 
is defined by the formula
$$
l_Z := \int_0^1 \frac{| \dot{Z}(t) |}{Y(t)}\,{\rm d} t
$$
where the dot indicates differentiation with respect to the parameter $t$.
This length defines a hyperbolic metric 
$\varrho (z_0,z_1) : \,{\mathbb H} \times {\mathbb H} \rightarrow {\mathbb R}_+$ via
$$
\varrho(z_0,z_1) := \min_{\left \{ Z :\, Z(0)=z_0,\;Z(1)=z_1\right \}} l_Z\,.
$$

The hyperbolic distance between two points may be calculated explicitly; one finds
$$
\text{ch}  \left [ \varrho (z_0,z_1) \right ]  = \frac{(x_0-x_1)^2+ y_0^2 + y_1^2}{2 y_0 y_1}\,.
$$

\begin{center}
\linethickness{2mm}
\color{light-grey}{\line(1,0){360}}
\end{center}
\begin{exercise}
Show that, for every $z \in {\mathbb H}$,
$$
\text{th} \left [ \frac{\varrho(z,{\rm i})}{2} \right ] = \left | \frac{z-{\rm i}}{z+{\rm i}} \right |
$$
\label{hyperbolicDistanceExercise1}
\end{exercise}
\begin{center}
\linethickness{2mm}
\color{light-grey}{\line(1,0){360}}
\end{center}

\subsection{Decay of the transmission coefficient}
\label{decaySubsection}
Let us now return to our impurity model. Formula (\ref{riccatiScattering}) for the reflexion coefficient is in terms of linear fractional
transformations with {\em real} coefficients. Such transformations play a very important part in the analysis of Poincar\'e's 
half-plane model of hyperbolic geometry.

\begin{center}
\linethickness{2mm}
\color{light-grey}{\line(1,0){360}}
\end{center}
\begin{exercise}
Let $A \in \text{SL} (2,{\,\mathbb R})$ and $Z : [0,1] \rightarrow {\mathbb H}$. Consider the transformed path
$$
{\mathcal A} (Z) := [0,1] \rightarrow {\mathcal A}(Z(t))\,.
$$
Show that
$$
l_{{\mathcal A}(Z)} = l_Z\,.
$$
In words, the length functional is invariant under the linear fractional transformation ${\mathcal A}$.
\label{hyperbolicLengthExercise}
\end{exercise}
\begin{center}
\linethickness{2mm}
\color{light-grey}{\line(1,0){360}}
\end{center}

It follows from this exercise that the Poincar\'{e} metric $\varrho$
is also invariant under the linear fractional transformations associated with $\text{SL}(2,{\mathbb R})$, i.e.
$$
\varrho \left ( {\mathcal A} (z_0), {\mathcal A}(z_1) \right ) = \varrho(z_0,z_1) \quad \text{for every $A \in \text{SL}(2,{\mathbb R})$}\,.
$$

\begin{center}
\linethickness{2mm}
\color{light-grey}{\line(1,0){360}}
\end{center}
\begin{exercise}
Show that, for every 
$$
A = \begin{pmatrix}
a & b \\
c & d
\end{pmatrix}
\in \text{SL} (2,{\mathbb R})\,,
$$
there holds
$$
2\, \text{ch} \left [ \varrho \left ({\rm i} ,{\mathcal A}({\rm i}) \right )  \right ] = | A |^2 := a^2+b^2 +c^2+d^2\,.
$$
\label{hyperbolicDistanceExercise2}
\end{exercise}
\begin{center}
\linethickness{2mm}
\color{light-grey}{\line(1,0){360}}
\end{center}

We now have in place all the elements needed in order to relate the transmission coefficient to the the product of matrices: By combining the result of Exercise \ref{hyperbolicDistanceExercise1}  with Formula (\ref{reflexionFormula}), we deduce
$$
\left | R_{L} \right | = \text{th} \left [  \frac{\varrho \left ({\mathcal X}_{n+1}({\rm i}),{\rm i}\right )}{2} \right ]
$$
and hence, by the result of Exercise \ref{reflexionTransmissionExercise},
$$
\left | T_{L}  \right |^2 = \frac{1}{\text{ch}^2 \left [ \frac{\varrho \left ( {\mathcal X}_{n+1}({\rm i}),{\rm i}\right )}{2} \right ]} = \frac{2}{1+\text{ch} \left [  \varrho \,\left ( {\mathcal X}_{n+1}({\rm i}),{\rm i} \right ) \right ]}\,.
$$
Furthermore, by using Formula (\ref{riccatiScattering}) and the result of Exercise \ref{hyperbolicDistanceExercise2}, we obtain
\begin{equation}
\left | T_{L}  \right |^2 = \frac{4}{2+\left | \Pi_n \right |^2}\,.
\label{transmissionFormula}
\end{equation}
In particular, if the Lyapunov exponent is positive, then 
\begin{equation}
\lim_{L \rightarrow \infty} \frac{\ln \left | T_L  \right |}{L} = - \gamma (k^2)\,.
\label{transmissionDecay}
\end{equation}
This says that the coefficient of transmission through a disordered sample
{\em decays exponentially with the sample length} and that the decay rate is precisely the Lyapunov exponent.

\subsection{Distribution of the reflexion phase}
\label{phaseSubsection}
It follows in particular that a plane wave incident on an infinite disordered sample is totally reflected: 
$$
\lim_{L \rightarrow \infty} \left | R_L \right | = 1\,.
$$
The distribution of the phase of $R_L$ for large samples is an interesting observable 
which has received some attention in the physics literature (Barnes and Luck 1990;  Sulem 1973).  

\begin{figure}[htbp]
\vspace{7.5cm} 
\includegraphics{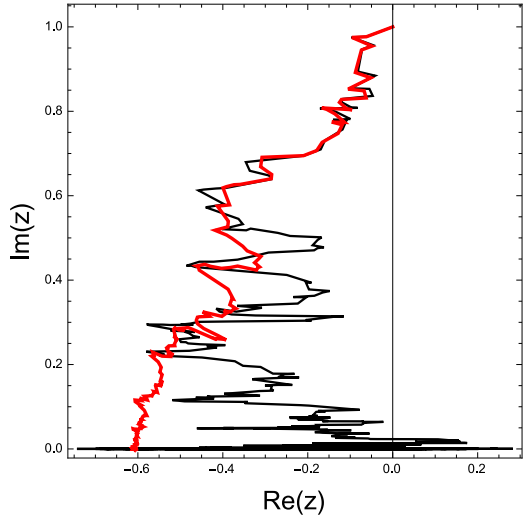}
\caption{The red curve corresponds to a typical trajectory of the backward iteration 
$\{ {\mathcal X}_n \}_{n \in {\mathbb N}}$
defined by (\ref{hyperbolicContinuedFraction}).
The black curve shows the trajectory of the corresponding forward iteration.} 
\label{forwardBackwardFigure} 
\end{figure}

For our impurity models,
Equation (\ref{reflexionFormula}) expresses the reflexion coefficient for a sample
of length $L = x_{n+1}$ in terms of the finite continued
fraction
\begin{equation}
{\mathcal X}_{n+1} ({\rm i}) = {\mathcal A}_0^{-1} \circ {\mathcal A}_1^{-1} \circ \cdots \circ {\mathcal A}_n^{-1} ({\rm i})\,.
\label{hyperbolicContinuedFraction}
\end{equation}
For every $n$, its value is a point in the half-plane ${\mathbb H}$. However, since $\lambda>0$,
the matrices $A_n$ are real, and so, if the continued fraction converges, its limit must be a {\em real}
random variable, say
\begin{equation}
X_\infty := \lim_{n \rightarrow \infty} {\mathcal X}_n ({\rm i}) \in {\mathbb R}\,.
\label{infiniteContinuedFraction}
\end{equation}
This is illustrated in Figure \ref{forwardBackwardFigure}.
This limit plays for the scattering problem the same part as the Weyl coefficient $w(\lambda)$ for the spectral
problem. In particular, the probability density of $-X_\infty$ is the stationary probability density $f$ of the Riccati process. Thus, if we write
$$
\lim_{L \rightarrow \infty} R_L = {\rm e}^{{\rm i} \Theta_\infty}
$$
then the probability density of the reflexion phase $\Theta_{\infty}$ of an infinite sample is
$$
\frac{1}{1+\cos \theta} \,f \left ( \frac{\sin \theta}{1+\cos \theta} \right )\,.
$$
Explicit formulae for the density function $f$ for the Frisch--Lloyd 
model of Example \ref{nieuwenhuizenExample} and for  the Dyson model
of Exercise \ref{firstLetacExercise} may be found in Comtet {\em et al.} (2010)
and Marklof {\em et al.} (2008) respectively.

The fact that a limiting distribution does exist is not entirely trivial.
Although ${\mathcal X}_{n+1}$ is like the continued fraction for the Weyl coefficient of the truncated spectral problem that we studied in \S \ref{spectralSection}, there is one significant difference between the two: for the scattering problem,
$\lambda = k^2=1$, and so
{\em the point at infinity is no longer of the limit-point type} . This means that the Weyl--Titchmarsh theory cannot be invoked to deduce the convergence of ${\mathcal X}_{n+1}$ as $n \rightarrow \infty$. A different approach is required, and
Bougerol and Lacroix (1985), Chapter 2, show that convergence takes place if the Lyapunov exponent is strictly positive.
Furstenberg's theorem, which will be discussed next, gives explicit conditions on the distribution of the $A_n$ for this to be the case, and so the scattering problem underlines the practical significance of Furstenberg's theory.

\section{The Lyapunov exponent of a product of random matrices}
\label{lyapunovExponentSection}
We now leave the subject of impurity models and turn our attention to
the large-$n$ behaviour of a the product
\begin{equation}
\Pi_n := A_n \cdots A_2 A_1
\label{productOfMatricesAgain}
\end{equation}
when the $A_n$ are arbitrary random $d \times d$ invertible matrices, i.e. $A_n \in \text{GL} (d,{\mathbb R})$. 
More precisely, we shall be concerned with the  limit
$$
\lim_{n \rightarrow \infty} \frac{1}{n} \ln \left | \Pi_n {\mathbf x} \right |
$$
where ${\mathbf x}$ is some non-random, non-zero vector and $| \cdot |$ is the familiar Euclidean norm in ${\mathbb R}^d$, i.e.
\begin{equation}
| {\mathbf x} | := \left ( \sum_{j=1}^d x_j^2 \right )^{\frac{1}{2}}\,.
\label{vectorEuclideanNorm}
\end{equation}
Also, we shall restrict our attention to the case where
the $A_n$ are {\em independent and identically distributed}. We denote by $\mu$ the distribution from which they are drawn, i.e.
for every set $S$ of matrices,
$$
{\mathbb P} \left ( A_n \in S \right ) = \int_S {\rm d} \mu (A)\,.
$$

In order to get a feel for the
problem, let us begin by considering some simple particular cases.
\begin{example}
When $d=1$, the $A_n$ are just {\em numbers}. Then $| \cdot |$ is
the absolute value and we have
$$
\frac{1}{n} \ln \left | \Pi_n x \right |=
\frac{1}{n} \sum_{j=1}^n \ln |A_j| + \frac{\ln |x|}{n} \xrightarrow[n \rightarrow \infty]{\text{a.s.}} {\mathbb E} \left ( \ln | A | \right )
$$
where
$A$ is $\mu$-distributed. This result is nothing but the familiar {\em Law of Large Numbers}, which says that,
if one draws numbers repeatedly and independently from the same distribution, then the average value 
after many draws approaches the mean of that distribution.
In the $1 \times 1$ case, the growth of the product is therefore easily expressed in terms
of the distribution $\mu$, since
$$
{\mathbb E} \left ( \ln | A | \right ) = \int_{\text{GL} (1,{\mathbb R})} \ln |A| \, {\rm d} \mu (A)\,.
$$
This argument extends to 
the case where $d$ is arbitrary but the $A_n$ are {\em diagonalisable} and {\em commute}, i.e.
$$
A_m A_n = A_n A_m \;\;\text{for every $m,n \in {\mathbb N}$}\,.
$$
Then, for every realisation of the sequence $\{ A_n \}_{n \in {\mathbb N}}$, the matrices $A_n$ share the same eigenvectors. So one can find a an invertible matrix independent of $n$, say $M$, such that
for every $n \in {\mathbb N}$
$$
M A_n M^{-1} = \text{diag} \left ( \lambda_{n,j} \right )
$$
where the $\lambda_{n,j}$, $1 \le j \le d$, are the eigenvalues of $A_n$.
\label{numbersExample}
\end{example}

Denote by $| A |$ the norm of the matrix $A$, i.e.
\begin{equation}
|A | := \sup_{|{\mathbf x}| \le 1} \left | A {\mathbf x} \right |\,.
\label{euclideanNorm}
\end{equation}

For typical $A$, $B \in \text{GL} \left ( d, {\mathbb R} \right )$, we have
$$
\left | A B \right | \le | A | \,|B |
$$
rather than the strict equality of the $d=1$ case. So, unless the matrices in the product commute, the argument used in Example \ref{numbersExample} breaks down. 
Nevertheless, as we shall see, 
the result
$$
\frac{1}{n} \ln \left | \Pi_n {\mathbf x} \right | \xrightarrow[n \rightarrow \infty]{\text{a.s.}} \gamma_{\mu}
$$
holds under conditions that are quite natural. The number $\gamma_{\mu}$ is called the {\em Lyapunov exponent} of the product.

\begin{remark}
The norm (\ref{euclideanNorm}) induced by the choice (\ref{vectorEuclideanNorm}) differs from that used in \S \ref{scatteringSection}, Exercise \ref{hyperbolicDistanceExercise2}.
We point out, however, that in a finite-dimensional space, all norms are equivalent. Hence the value of the Lyapunov exponent does not depend on the particular norm chosen; the choice is merely a matter of convenience.
\label{normRemark}
\end{remark}

Let ${\mathbf x}$ be a non-zero vector and set
\begin{equation}
\notag
Y_1= \ln \left |  A_1 \frac{{\mathbf x} }{| \mathbf x |} \right |\,, \;
Y_2= \ln \left | A_2 \frac{{A_1\mathbf x} }{|A_1 \mathbf x |} \right |\,, 
\ldots \,,\;Y_n= \ln \left | A_n \frac{{A_{n-1}\cdots A_1\mathbf x} }{|A_{n-1}\cdots A_1 \mathbf x |} \right |\,.
\end{equation} 
Then we have the following ``telescopic'' formula
\begin{equation}
\frac{1}{n}\ln \frac{\left |A_nA_{n-1}\cdots A_1\mathbf x \right |}{|\mathbf x|} =\frac{1}{n} \sum_{j=1}^n Y_j\,.
\label{telescopicProperty}
\end{equation}

\subsection{The Cohen--Newman example}
\label{cohenNewmanSubsection}
Cohen \& Newman (1984) remarked the following: suppose that $\mu$, the distribution of $A$, is such that
the random variable $\left | A {\mathbf u} \right |$
has the same distribution for every unit vector $\mathbf u$. 
Then the random variables $Y_j$ defined above are identically distributed. They are also {\em independent}.
Indeed, if the $S_j$ are measurable sets, then
\begin{multline}
\notag
{\mathbb P} \left ( Y_{n+1} \in S \,\Bigl |  \;Y_j \in S_j, \,1 \le j \le n \right ) = \\
{\mathbb P} \left ( \ln \left | A_{n+1} \frac{{A_{n}\cdots A_1\mathbf x} }{|A_{n}\cdots A_1 \mathbf x |} \right | \in S \,\Bigl |  \;Y_j \in S_j, \,1 \le j \le n \right ) \\
= {\mathbb P} \left ( \ln \left | A_{n+1} {\mathbf u} \right | \in S \,\Bigl |  \;Y_j \in S_j, \,1 \le j \le n \right ) 
= {\mathbb P} \left ( \ln \left | A_{n+1} {\mathbf u} \right | \in S \right )
\end{multline}
where ${\mathbf u}$ is any fixed non-random vector. Hence
$$
{\mathbb P} \left ( Y_{n+1} \in S \,\Bigl |  \;Y_j \in S_j, \,1 \le j \le n \right ) = {\mathbb P} \left ( Y_{n+1} \in S \right )\,.
$$

We may therefore use the Law of Large Numbers to conclude from Equation (\ref{telescopicProperty}) that
$$
\gamma_{\mu} =  {\mathbb E} \left ( \ln \left | A \frac{{\mathbf x}}{\left | {\mathbf x} \right |} \right | \right )\,.
$$

\begin{center}
\linethickness{2mm}
\color{light-grey}{\line(1,0){360}}
\end{center}
\begin{exercise}
Let 
$$
A := 
\begin{pmatrix}
\alpha & \beta \\
0 & \frac{1}{\alpha}
\end{pmatrix}  
\begin{pmatrix}
\cos \theta & -\sin \theta \\
\sin \theta & \cos \theta
\end{pmatrix}
\,.
$$
where $\alpha \ne 0$ and $\beta$ are fixed, non-random numbers,
and $\theta$ is uniformly distributed over $[0,2\pi)$. 

\begin{enumerate}[label=(\alph*)]
\item Show that
$$
\left | A {\mathbf u} \right | 
$$
has the same law for every unit vector ${\mathbf u}$.

\item Deduce that the Lyapunov exponent for the corresponding product is
$$
\gamma_{\mu} = \ln\frac{{(\alpha+1/\alpha})^2+\beta^2}{4}\,.
$$
\end{enumerate}
\label{alainExercise1}
\end{exercise}
\begin{center}
\linethickness{2mm}
\color{light-grey}{\line(1,0){360}}
\end{center}

\subsection{Application to the Frisch--Lloyd model}
\label{alainExampleSubsection}

Let us apply the result of the previous exercise to a disordered system.
Consider the Frisch--Lloyd model with impurities that are uniformly distributed in $(0,\infty)$ with mean spacing $\ell$,
but {\em fixed deterministic} coupling constants. In other words, for every $j \in {\mathbb N}$,
$$
{\mathbb P} ( \ell_{j-1} \in (a,b)) =  \int_a^b {\rm d} x\, \frac{1}{\ell} {\rm e}^{-x/\ell} \quad \text{and} \quad v_j = v\,.
$$

\begin{center}
\linethickness{2mm}
\color{light-grey}{\line(1,0){360}}
\end{center}
\begin{exercise}
With these assumptions:

\begin{enumerate}[label=(\alph*)]
\item Show that, if one uses
$$
\begin{pmatrix}
\psi'(x-) \\
k \psi (x)
\end{pmatrix}
\;\;\text{instead of}\;\;
\begin{pmatrix}
\psi'(x-) \\
\psi (x)
\end{pmatrix}
$$
then the solution of the Frisch--Lloyd model may be expressed as
$$
\begin{pmatrix}
\psi'(x_{n+1}-) \\
k \psi (x_{n+1})
\end{pmatrix}
= \begin{pmatrix}
\cos(k \ell_n) & -\sin (k \ell_n) \\
\sin (k \ell_n) & \cos (k \ell_n)
\end{pmatrix}
\Pi_n
\begin{pmatrix}
\psi'(0-) \\
k \psi (0)
\end{pmatrix}
$$
where
$$
\Pi_n := A_n \cdots A_1 
$$
with
$$
A_j := 
\begin{pmatrix}
1 & v/k \\
0 & 1
\end{pmatrix}
\begin{pmatrix}
\cos(k \ell_{j-1}) & -\sin (k \ell_{j-1}) \\
\sin (k \ell_{j-1}) & \cos (k \ell_{j-1})
\end{pmatrix}
\,.
$$

\item Introduce the reduced phase
$$
\theta_j := 2 \pi \left \{ \frac{k \ell_{j-1}}{2 \pi} \right \} \in [0,2 \pi)\,.
$$
Equivalently,
$$
\theta_j = k \ell_{j-1} \;\;\text{mod} \,2 \pi\,.
$$
Show that, for large $k \ell$--- that is, when the mean spacing between successive impurities is large compared to the wavelength---
$\theta_j$ is approximately uniformly distributed on $[0,2 \pi)$.
\item Use the result of Exercise \ref{alainExercise1} to find the limit
$$
\lim_{L \rightarrow \infty} \frac{\ln \sqrt{\left | \psi'(L-,k^2) \right |^2 + k^2 \left | \psi(L,k^2) \right |^2}}{L}\,.
$$
Compare this with the result obtained by Bienaim\'e \& Texier (2008).
\end{enumerate}
\label{alainExercise2}
\end{exercise}
\begin{center}
\linethickness{2mm}
\color{light-grey}{\line(1,0){360}}
\end{center}

\section{Furstenberg's formula for the Lyapunov exponent}
\label{furstenbergFormulaSection}

We stress that the telescopic property (\ref{telescopicProperty}) always holds but that, in general the $Y_j$ are 
{\em neither  independent  nor identically-distributed}
random variables. So the Law of Large Numbers cannot be used.
Instead, as we shall see in this section, the Lyapunov exponent can be expressed, via the ergodic theorem, as a  functional of a certain Markov chain. This fact will lead to a general formula for the Lyapunov exponent
which is, in some sense, the ``product version'' of the usual  Law of Large Numbers.
Importantly, the formula involves  averaging--- not only over the matrices in the product--- but also  over a certain space  called  the {\em projective space}.

\subsection{The projective space}
\label{projectiveSubsection}
We say that two vectors ${\mathbf x}$ and ${\mathbf y}$ in ${\mathbb R}^d$ have the same {\em direction} if one is a scalar multiple of the
other, i.e. there exists $c \in {\mathbb R} \backslash \{0\}$ such that
$$
{\mathbf y} = c {\mathbf x}\,.
$$
This defines an equivalence relation in ${\mathbb R}^d$; the set of all directions can be partitioned into equivalence classes,
and each equivalence class can be identified
with a straight line through the origin. The set of all such lines is called the {\em projective space} and is denoted $P \left ( {\mathbb R}^d \right )$.

\begin{notation}
We shall use $\overline{\mathbf x}$ to denote the direction
of the vector ${\mathbf x} \in {\mathbb R}^d \backslash \{ {\mathbf 0} \}$. We shall frequently abuse this notation
by treating the map
$$
{\mathbf x} \mapsto \overline{\mathbf x}
$$
as though it were invertible. That is, we shall
use $\overline{\mathbf x}$ to denote also an arbitrary element of $P \left ( {\mathbb R}^d \right )$ to which we then
associate a vector ${\mathbf x} \in {\mathbb R}^d \backslash \{ {\mathbf 0} \}$ whose direction is
$\overline{\mathbf x}$. This does no harm, as long as the result does not depend on the choice of the particular vector ${\mathbf x}$.
\end{notation}

\subsection{A Markov chain and its stationary distribution}
\label{markovSubsection}
Let $A \in \text{SL} \left ( d, {\mathbb R} \right )$ and $\overline{\mathbf x} \in P({\mathbb R}^d)$.
Since, by assumption, 
the determinant of $A$ does not vanish, the vector $A {\mathbf x}$ is non-zero and so lies along a line in ${\mathbb R}^d$.
\begin{notation}
$$
A \cdot \overline{\mathbf x} := \overline{A {\mathbf x}} \,.
$$
\end{notation}

\begin{center}
\linethickness{2mm}
\color{light-grey}{\line(1,0){360}}
\end{center}
\begin{exercise}
Show that, for every $A_1$, $A_2 \in \text{SL} (d,{\mathbb R})$ and every 
$\overline{\mathbf x} \in P({\mathbb R}^d)$,
$$
A_1 A_2 \cdot \overline{\mathbf x}  = A_1 \cdot  \left ( A_2 \cdot \overline{\mathbf x} \right )\,.
$$
One says that the group $\text{SL}(d,{\mathbb R})$ {\em acts} on the projective space.
\label{actionExercise}
\end{exercise}
\begin{center}
\linethickness{2mm}
\color{light-grey}{\line(1,0){360}}
\end{center}

Suppose now that the $A_n$ are independent and $\mu$-distributed. Let $\overline{\mathbf x} \in P({\mathbb R}^d)$ and define the random sequence $\{ \overline{\mathbf x}_n \}_{n \in {\mathbb N}}$ by recurrence: 
$$
\overline{\mathbf x}_1 := \overline{\mathbf x} \;\;\text{and}\;\; \overline{\mathbf x}_{j+1} := A_j \cdot \overline{\mathbf x}_j
\;\;\text{for $j=1,2,\cdots$}
$$
It is easy to see that the sequence
$$
\left \{ \left ( A_n,\,\overline{\mathbf x}_n \right ) \right \}_{n \in {\mathbf N}}
$$
is a Markov chain in the product space $\text{SL} (d,{\mathbb R}) \times P \left ( {\mathbb R}^d \right )$.
Furthermore, the telescopic formula (\ref{telescopicProperty}) 
may be expressed as 
\begin{equation}
\frac{1}{n}\ln \frac{\left |A_nA_{n-1}\cdots A_1\mathbf x \right |}{|\mathbf x|} =
\frac{1}{n} \sum_{j=1}^n F \left ( A_j, \overline{\mathbf x}_j \right ) 
\label{telescopicProperty2}
\end{equation}
where $\overline{\mathbf x}$ is the direction of ${\mathbf x}$ and
$$
F \left ( A,\overline{\mathbf x} \right )  = \ln \left | A \frac{\mathbf x}{| {\mathbf x} |} \right |\,.
$$

\begin{definition}
Let $\nu$ be a probability distribution on $P \left ({\mathbb R}^d \right )$.
We say that $\nu$ is {\em stationary for $\mu$} if, for every $\nu$-distributed direction
$\overline{\mathbf x}$ and every independent $\mu$-distributed matrix $A$, we have
$$
A \cdot \overline{\mathbf x} \overset{\text{law}}{=} \overline{\mathbf x}\,.
$$
\label{stationarityDefinition}
\end{definition}

Now, suppose that there exists a unique
$\mu$-stationary probability measure $\nu$ on $P \left ( {\mathbb R}^d \right )$ and that $\overline{\mathbf x}$
is $\nu$-distributed. It is easy to see from the construction of the Markov chain that
$$
\left ( A_n,\,\overline{\mathbf x}_n \right ) \overset{\text{law}}{=} \left ( A_1,\,\overline{\mathbf x} \right )
$$
for every $n \in {\mathbb N}$. In other words, the product measure $\mu (A) \nu (\overline{\mathbf x})$
is the stationary distribution of the Markov chain. By the ergodic theorem, we therefore deduce from
Equation (\ref{telescopicProperty2}) that
$$
\frac{1}{n}\ln \frac{\left |A_nA_{n-1}\cdots A_1\mathbf x \right |}{|\mathbf x|} \xrightarrow[n \rightarrow \infty]{}
\int_{\text{SL} \left ( d,{\mathbb R} \right )} {\rm d} \mu (A) \int_{P \left ({\mathbb R}^d  \right )}
{\rm d} \nu ( \overline{\mathbf x} ) \, F(A,\overline{\mathbf x})\,.
$$
In other words,
\begin{equation}
\gamma_{\mu} = 
\int_{\text{SL} \left ( d,{\mathbb R} \right )} {\rm d} \mu (A) \int_{P \left ({\mathbb R}^d  \right )}
{\rm d} \nu ( \overline{\mathbf x} )\,
\ln \frac{\left | A {\mathbf x} \right |}{\left | {\mathbf x} \right |}\,.
\label{furstenbergFormula}
\end{equation}

This is Furstenberg's formula for the Lyapunov exponent of the product of random matrices.
The essential
difficulty in its practical application is that the $\mu$-stationary measure $\nu$ is not known a priori.

\subsection{The case $d=2$}
\label{twoByTwoSubsection}
In this case, we can make the foregoing discussion much more concrete but relating it to familiar
geometrical concepts.
We speak of $P({\mathbb R}^2)$ as the {\em projective line}. To specify a particular member
of the projective line, we may use (the reciprocal of) its slope:
\begin{equation}
{\mathbf 0} \ne {\mathbf x} = \begin{pmatrix} x_1 \\ x_2 \end{pmatrix} \implies \overline{{\mathbf x}} = z := x_1/x_2\,.
\label{furstenbergRiccatiVariable}
\end{equation}
The direction $z$  may be finite or infinite.
This defines a bijection between the projective line
and the set of ``numbers'' $\overline{\mathbb R} = {\mathbb R} \cup \{\infty\}$. With some abuse of notation, we shall sometimes
write
$$
P \left ( {\mathbb R}^2 \right ) = \overline{\mathbb R}\,.
$$
The reader will immediately recognise that this variable $z$ defined by 
Equation (\ref{furstenbergRiccatiVariable}) coincides with the Riccati variable introduced in the particular context
of impurity models. The relevance of the projective space in the more general context may be understood intuitively as follows.  Suppose for simplicity that all the matrices have
{\em positive} determinant, so that there is no further loss of generality in assuming that their determinant is one.
When  $d=2$, we can  write the product in the column form
$$
\prod_{j=1}^n A_j = \begin{pmatrix} {\mathbf p}_n & {\mathbf q}_n \end{pmatrix}\,.
$$
Recall the geometrical interpretation 
of the determinant in the $2 \times 2$
case: its modulus is the area of the parallelogram spanned by the columns. We see that
unimodularity implies that
$$
| {\mathbf p}_n | \,| {\mathbf q}_n | | \sin \theta_n | = 1
$$
where $\theta_n$ is the {\em angle} between the columns.
Hence, if we show that the columns tend to align along the same direction, then $\theta_n \rightarrow 0$ as $n \rightarrow \infty$
and at least one of $| {\mathbf p}_n |$ or $| {\mathbf q}_n |$ must grow.

Let
$$
A := \begin{pmatrix}
a & b \\
c & d
\end{pmatrix}
\,.
$$
and
$$
{\mathbf x} = \begin{pmatrix} x_1 \\ x_2 \end{pmatrix} \ne {\mathbf 0}\,.
$$ 

The reciprocal of the slope of this line is
$$
\frac{a x_1 + b x_2}{c x_1 + d x_2} = \frac{a z + b}{c z + d} = {\mathcal A}(z)
$$
where ${\mathcal A}$ is the linear fractional transformation introduced in Equation (\ref{linearFractionalTransformation}).
The matrix $A$ has ``acted'' on the line of direction of $z = x_1/x_2$ and mapped it to another
line whose direction is
$$
A \cdot z = {\mathcal A}  (z) 
$$

Returning to the formula (\ref{furstenbergFormula})  for the Lyapunov exponent,
this is how one should read the right-hand side when $d=2$: the
number
$$
\frac{ | A {\mathbf x} |}{| \mathbf x |} = \left | A \frac{{\mathbf x}}{|{\mathbf x}|} \right | 
$$
depends only on $A$ and on the {\em direction} $z \in \overline{\mathbb R}$ of the non-zero vector ${\mathbf x}$. So we can write
$$
\frac{ | A {\mathbf x} |}{| \mathbf x |} = \frac{ \left | A \begin{pmatrix}  z \\ 1 \end{pmatrix} \right |}{\left | \begin{pmatrix} z \\ 1
 \end{pmatrix} \right |}\,.
$$
The formula for the Lyapunov exponent then takes the more readable form
\begin{equation}
\gamma_{\mu} = \int_{\overline{{\mathbb R}}} \int_{\text{SL} \left (2,{\mathbb R} \right )} \ln \frac{\left | A \begin{pmatrix} z \\ 1
\end{pmatrix} \right |}{ \left |  
\begin{pmatrix} z \\ 1 \end{pmatrix} \right |} {\rm d} \mu (A) \,{\rm d} \nu ( z )\,.
\label{twoByTwoFurstenbergFormula}
\end{equation}

How can one find $\nu$? 
In the case $d=2$, if we assume that $\nu$ has a density $f$, then this density satisfies the Dyson--Schmidt equation (\ref{letacEquation}):
$$
f(z) = {\mathbb E} \left ( \left [ f \circ {\mathcal A}^{-1} \right ] (z) \frac{{\rm d} {\mathcal A}^{-1}}{{\rm d} z} (z) \right )\,.
$$
The existence and uniqueness of a solution is intimately connected with the convergence of the continued fraction
\begin{equation}
\lim_{n \rightarrow \infty} {\mathcal A}_1 \circ \cdots \circ {\mathcal A}_n (z)\,.
\label{backwardIteration}
\end{equation}
In particular, for the impurity models,
we saw in \S \ref{dysonSchmidtSection} that, under quite mild assumptions, the convergence of this continued fraction followed from the fact that the differential problem was in the limit-point case.
We also saw that, when $f$ exists, for $\lambda \in {\mathbb R}$,
$$
\gamma (\lambda+{\rm i}0)= \text{Re} \,\Omega (\lambda+{\rm i}0) 
= \dashint_{\mathbb R} z f(z)\,{\rm d} z\,.
$$
On the other hand,
$$
\gamma (\lambda+{\rm i}0) = \left ( \lim_{n \rightarrow \infty} \frac{n}{x_n} \right )\,\gamma_{\mu}
$$
where $\gamma_{\mu}$ is the Lyapunov exponent--- in the sense of Furstenberg---
of the product of matrices, evaluated
at $\lambda+{\rm i}0$,
associated with the impurity model.
It is striking that Formula (\ref{twoByTwoFurstenbergFormula}) for $\gamma_{\mu}$
looks much more complicated than the formula for $\gamma(\lambda+{\rm i} 0)$. 
The fact the {\em both formulae are correct} is proved in
Comtet {\em et al.} (2010). We shall not reproduce the proof here, but some of the calculations
in later sections will illustrate the simplifications that can occur when applying Furstenberg's formula
to specific products of $2 \times 2$ matrices.

\section{Furstenberg's Theorem}
\label{furstenbergSection}
We have so far emphasised the practical aspects of the {\em computation} of the 
Lyapunov exponent. It should be apparent by now that explicit calculations are possible only in 
exceptional cases. For theoretical purposes, it is often sufficient to determine whether or not the Lyapunov exponent is strictly positive. Furstenberg (1963) and others developed a general theory with the aim of addressing this question in the abstract context of random walks on groups.
Here we apply it to the case where the group is $G := \text{SL} \left (2 ,{\mathbb R} \right )$.
For the proofs of the results stated,
the reader should consult (Bougerol and Lacroix 1985; Carmona and Lacroix 1990).

Using the matrix norm introduced in \S \ref{lyapunovExponentSection}, we can speak of limits of sequences, and of bounded and closed sets in $G$.

\begin{notation}
Given a probability measure $\mu$ on $G$, we denote by $G_{\mu}$ the smallest closed subgroup of $G$ containing the support
of $\mu$.
\label{GmuNotation}
\end{notation}

\begin{definition}
A subgroup $H$ of $G$ is said to be {\em strongly irreducible} if there is no finite union
$$
S = \cup_{j=1}^m S_j
$$ 
of one-dimensional subspaces $S_j$ of ${\mathbb R}^2$ such that, for every $A \in H$,
$$
A(S) = S\,.
$$ 

By extension, we say that the measure $\mu$ itself is strongly irreducible if $G_{\mu}$ is strongly irreducible.
\label{irreducibilityDefinition}
\end{definition}

The following criterion will be useful.

\begin{proposition}
If $G_{\mu}$ is unbounded, then $G_{\mu}$ is strongly irreducible if and only if the following holds:
for every $z \in P \left ( {\mathbb R}^2 \right )$,
$$
\# \left \{ {\mathcal A} ( z) \,:\; A \in G_{\mu} \right \} > 2\,.
$$
\label{irreducibilityCriterion}
\end{proposition}

\begin{remark}
Intuitively, in terms of the impurity models, the concept of strong irreducibility expresses the requirement
that there should should be ``enough'' disorder.
\label{irreducibilityRemark}
\end{remark}

\begin{theorem}[Furstenberg]
Suppose that
$$
{\mathbb E} \left ( \ln | A | \right ) < \infty\,.
$$
Suppose also that $G_\mu$ is strongly irreducible and unbounded. Then the following statements are true:
\begin{enumerate}
\item There exists a unique measure $\nu$ on $P \left ({\mathbb R}^2 \right )$
that is stationary for $\mu$. 
\item For every non-random, non-zero vector
${\mathbf x}$, we have
$$
\frac{1}{n} \ln \left | \Pi_n {\mathbf x} \right | \xrightarrow[n \rightarrow \infty]{\text{a.s.}} \gamma_{\mu}
$$
where $\gamma_{\mu}$ is given by Formula (\ref{twoByTwoFurstenbergFormula}).
\item $\gamma_{\mu} > 0$.
\end{enumerate}
\label{furstenbergTheorem}
\end{theorem}


We shall illustrate the use and the content of Furstenberg's Theorem by means of elementary examples. Some of these examples have in common that the product of matrices
arises from the solution of the difference equation
\begin{equation}
\psi_{n+1} = a_n \psi_{n} + b_n \psi_{n-1} \,, \quad n \in {\mathbb N}\,.
\label{differenceEquation}
\end{equation} 
This may be viewed as a generalisation of the Anderson model mentioned in \S \ref{andersonSubsection}.
Obviously, its general solution is given by 
$$
\begin{pmatrix}
\psi_{n+1} \\ \psi_n
\end{pmatrix}
= \Pi_n \begin{pmatrix}
\psi_{1} \\ \psi_0
\end{pmatrix}
$$
where the $A_j$ in the product (\ref{productOfMatrices}) are given by
$$
A_j = \begin{pmatrix}
a_j & b_j \\
1 & 0
\end{pmatrix}\,.
$$

\begin{example}
\label{FibonacciExample}
Our first example is deterministic! The Fibonacci sequence 
satisfies the recurrence relation
\begin{equation}
\psi_{n+1} = \psi_n + \psi_{n-1}\,, \;\; n \in {\mathbb N}\,,
\label{fibonacciRecurrence}
\end{equation}
with $\psi_0=\psi_1=1$. Alternatively,
$$
\begin{pmatrix}
\psi_{n+1} \\
\psi_n 
\end{pmatrix}
= \Pi_n \begin{pmatrix}
1 \\
1
\end{pmatrix}
$$
where
$$
A_j =  A := \begin{pmatrix}
1 & 1 \\
1 & 0
\end{pmatrix} \quad \text{for $j \in {\mathbb N}$}\,.
$$
We may think of $\Pi_n$ as a product of ``random'' matrices with a distribution $\mu$ whose mass is concentrated at $A$;
we express this as
$$
\mu = \delta_{A}\,.
$$

This $\mu$ is {\em not} strongly irreducible: the matrix $A$ has two eigenvectors, and the straight lines along which these
eigenvectors lie are invariant under multiplication by $A$. So Furstenberg's Theorem does not hold, and this manifests itself
in the fact that the limit of $Z_n(z)$ in Equation (\ref{backwardIteration}) depends on $z$. Indeed, let $z_\pm$ be the two roots
of 
$$
z = {\mathcal A}(z) = \frac{z+1}{z}\,.
$$
Then, for every $n \in {\mathbb N}$, we have
$$
Z_n (z_{\pm}) = z_{\pm}
$$
and so there are {\em two} invariant measures, namely $\nu_{+}$ and $\nu_-$, concentrated
respectively on $z_+$ and $z_-$.
\end{example}

\begin{example}
\label{randomFibonacciExample}
Consider the following randomised version of the previous example: Let the $A_j$ be drawn from
$$
\text{supp} \,{\mu} := \left \{
\begin{pmatrix}
-1 & 1 \\
1 & 0
\end{pmatrix} \,,\quad \begin{pmatrix}
1 & 1 \\
1 & 0
\end{pmatrix} 
\right \}
$$
with equal probability.

$G_{\mu}$ is larger than in the previous (deterministic) example, and it follows from Proposition \ref{irreducibilityCriterion} that Furstenberg's theorem holds.

The following illustrates the kind of manipulations involved in the calculation of the Lyapunov exponent: 
We write
$$
A = \begin{pmatrix}
a & 1 \\
1 & 0
\end{pmatrix}\,.
$$
Then
\begin{multline}
\notag
\gamma_{\mu} = \int_{\overline{\mathbb R}} \int_{\text{GL}(2,{\mathbb R})} \ln \frac{\left | A \begin{pmatrix} z \\ 1 \end{pmatrix} \right |}{\left | \begin{pmatrix} z \\ 1 \end{pmatrix} \right |}\,{\rm d} \mu (A) \,{\rm d} \nu(z) \\
= 
\frac{1}{2} \int_{\overline{\mathbb R}} \int_{\text{GL}(2,{\mathbb R})} \ln \frac{\left | A \begin{pmatrix} z \\ 1 \end{pmatrix} \right |^2}{\left | \begin{pmatrix} z \\ 1 \end{pmatrix} \right |^2}\,{\rm d} \mu (A) \,{\rm d} \nu(z) \\
= \frac{1}{2} \int_{\overline{\mathbb R}} \int_{\text{GL}(2,{\mathbb R})} \ln \frac{(a z + 1)^2 + z^2}{1+x^2}\,{\rm d} \mu (A) \,{\rm d} \nu(z) \\
= \frac{1}{2} \int_{\overline{\mathbb R}} \int_{\text{GL}(2,{\mathbb R})} \ln \left \{ \frac{z^2}{1+z^2} \left [ 1 +  \left [ {\mathcal A} ( z) \right ]^2 \right ]\right \}\,{\rm d} \mu (A) \,{\rm d} \nu(z) \\
= \frac{1}{2} \int_{\overline{\mathbb R}} \int_{\text{GL}(2,{\mathbb R})} \ln  \frac{z^2}{1+z^2} \,{\rm d} \mu (A) \,{\rm d} \nu(z)
\\
+ \frac{1}{2} \int_{\overline{\mathbb R}} \int_{\text{GL}(2,{\mathbb R})} \ln \left [ 1 +  \left [ {\mathcal A} ( z) \right ]^2 \right ]\,{\rm d} \mu (A) \,{\rm d} \nu(z)\,.
\end{multline}
At this point, we observe that, in the first of these integrals, the integrand is independent of $A$. Hence
$$
\frac{1}{2} \int_{\overline{\mathbb R}} \int_{\text{GL}(2,{\mathbb R})} \ln  \frac{z^2}{1+z^2} \,{\rm d} \mu (A) \,{\rm d} \nu(z) = \frac{1}{2} \int_{\overline{\mathbb R}} \ln  \frac{z^2}{1+z^2} \,{\rm d} \nu(x)\,.
$$ 
Furthermore, using the fact that $\nu$ is stationary for $\mu$,
$$
\frac{1}{2} \int_{\overline{\mathbb R}} \int_{\text{GL}(2,{\mathbb R})} \ln \left ( 1 +  \left [ {\mathcal A} (z) \right ]^2 \right )\,{\rm d} \mu (A) \,{\rm d} \nu(z)
= \frac{1}{2} \int_{\overline{\mathbb R}} \ln \left ( 1 +  z^2 \right ) \,{\rm d} \nu(z)\,.
$$
Putting these results together, we obtain a much deflated formula for the Lyapunov exponent:
$$
\gamma_{\mu} = \int_{\overline{\mathbb R}} \ln | z | \,{\rm d} \nu (z)\,.
$$

Viswanath (2000) considered and solved the problem of finding the $\mu$-stationary measure $\nu$ for this example.
It turns out that $\nu$ is not a smooth measure; it is {\em singular continuous}. There is no explicit formula
for it, but the measure of any real interval may be computed exactly to any desired accuracy by means of a recursion.
Then
$$
\gamma_\mu \in (0.1239755980, 0.1239755995 )\,.
$$
This value is, as it should be, smaller than the growth rate of the (deterministic) Fibonacci sequence, i.e.
$$
\ln \frac{\sqrt{5}+1}{2} = 0.481 \ldots
$$

\begin{figure}[htbp]
\vspace{8cm} 
\includegraphics{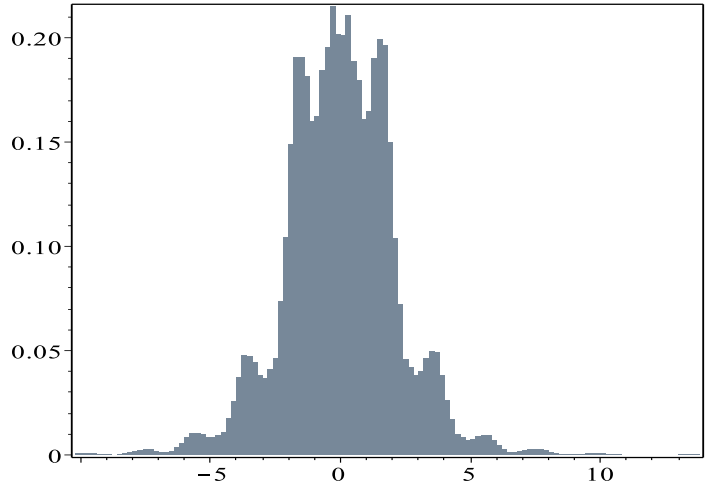}
\begin{picture}(0,0) 
\end{picture}  
\caption{A histogram of the first $8000$ of the forward iterates for the random Fibonacci example.}
\label{viswanathFigure} 
\end{figure}

Figure \ref{viswanathFigure} shows a histogram of the the first $8000$ terms of the
sequence defined by Equation (\ref{forwardIteration}) with a normal random variable as starting value.
The sequence is ergodic and so the histogram may be thought of as the ``graph'' of ``${\rm d} \nu (z)$''.
\end{example}

\begin{example}
\label{noGrowthExample}

Let us give an example, taken from (Bougerol and Lacroix 1985), of a product that does {\em not} grow.
Let $\alpha > 0$ and set
$$
D := \begin{pmatrix}
\alpha & 0 \\
0 & 1/\alpha
\end{pmatrix}
\;\;\text{and}\;\;
R := \begin{pmatrix}
0 & -1 \\
1 & 0
\end{pmatrix}\,.
$$
Consider the distribution $\mu$ supported on $\{ D,\,R\}$ such that
$$
{\mathbb P} ( A = D) = p \;\;\text{and}\;\; {\mathbb P} ( A = R) = 1-p \,, \;\; \text{where} \;p \in [0,1]\,.
$$
The matrix $R$ is a rotation matrix, i.e. $R {\mathbf x}$ is the vector obtained after rotating ${\mathbf x}$ by the angle $\pi/2$. It is therefore
obvious that, if $p=0$, then $\gamma_{\mu} = 0$. On the other hand,
if $p=1$ and $\alpha \ne 1$, one of the columns of the corresponding product of matrices will grow. What happens if $0 < p < 1$?

\begin{figure}[htbp]
\vspace{8cm} 
\includegraphics{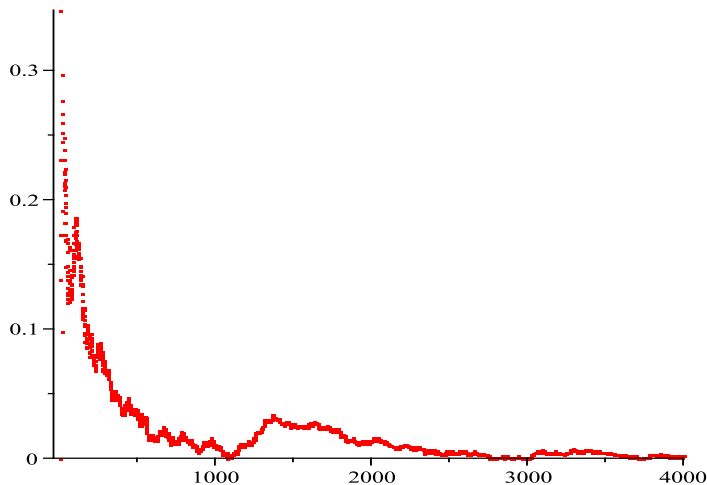}
\begin{picture}(0,0) 
\end{picture}  
\caption{Plot of $n^{-1} \ln \left | \prod_{j=1}^n A_j \right |$ against $n$ for the Bougerol--Lacroix  example.}
\label{bougerolFigure} 
\end{figure}

Figure \ref{bougerolFigure} shows a plot of  the quantity
$$
\frac{1}{n} \ln \left | \Pi_n \right |
$$
against $n$ in the case where $p=1/2$. The suggestion is that $\gamma_{\mu} = 0$. Turning to the statement of Furstenberg's theorem, we note that
$$
{\mathbb E} \left ( | A_1 | \right ) = p \ln \alpha\,. 
$$
For $0 < p < 1$, $G_{\mu}$ contains $D$, $R$, $D^{-1}$, $R^{-1}$ and every product of these. A little calculation shows
that
$$
G_{\mu} = \left \{ \begin{pmatrix} \beta^n & 0 \\ 0 & \beta^{-n} \end{pmatrix}\,, \;\begin{pmatrix} 0 & \beta^n \\ -\beta^{-n} & 0 \end{pmatrix} \;:
\; \beta \in \{ \pm \alpha, \pm 1/\alpha \},\, n \in {\mathbb Z}  \right \}\,.
$$
In particular, for every natural number
$n$, $D^n \in G_{\mu}$ and so $G_{\mu}$ is unbounded.

There only remains to examine the strong irreducibility assumption. Let $z = 0$. Then
we have
$$
\begin{pmatrix} \beta^n & 0 \\ 0 & \beta^{-n} \end{pmatrix} \cdot z = 0 \;\;\text{and}\;\;
\begin{pmatrix} 0 & \beta^n \\ -\beta^{-n} & 0 \end{pmatrix} \cdot z = \infty\,.
$$
We see that $\mu$ fails to satisfy the strong irreducibility criterion contained in Proposition \ref{irreducibilityCriterion}.
\end{example}

\end{document}